\documentclass[11pt,a4paper]{article}
\usepackage[lmargin=0.7in,rmargin=0.7in,tmargin=0.65in,bmargin=0.75in,footskip=0.25in]{geometry}
\usepackage[T1]{fontenc}
\usepackage{lmodern}
\usepackage[english]{babel}
\usepackage{amssymb,amsmath,amsthm}
\usepackage{xcolor}
\usepackage{graphicx}
\usepackage{tikz}
\usetikzlibrary{positioning,calc,arrows.meta}
\usepackage[colorlinks=true,allcolors=blue]{hyperref}

\newtheorem{theorem}{Theorem}[section]
\newtheorem{proposition}[theorem]{Proposition}
\newtheorem{corollary}[theorem]{Corollary}
\newtheorem{lemma}[theorem]{Lemma}
\newtheorem{definition}[theorem]{Definition}
\newtheorem{example}[theorem]{Example}
\newtheorem{remark}[theorem]{Remark}

\begin{document}

\begin{center}
{\large\textbf{Hidden Quantum Markov Processes:
Mealy Formulation and Observable Reduction}}

\bigskip

\textbf{ Soueidi El Gheteb}\\
Department of Mathematics, Minds- AI lab, Faculty of Sciences and Technologies,\\
University of Nouakchott, Nouakchott, Mauritania\\
\texttt{elkotobmedsalem@univ-nkc.mr}

\medskip

\textbf{Abdessatar Souissi}\\
Preparatory Institute for Scientific and Technical Studies La Marsa,
Carthage University, Tunisia\\
\texttt{abdessattar.souissi@ipest.rnu.tn}
\end{center}

\begin{abstract}
We formulate hidden quantum Markov processes in a Mealy
architecture using a single completely positive unital
transition expectation coupling the present hidden system,
the emitted output, and the next hidden system. Together
with a boundary state, it generates a quasi-local, finitely
correlated hidden--output state; for the commutative generator
associated with a classical Mealy kernel, this state is the full joint
law of the classical Mealy model.

Moore realizations correspond to transition expectations
that factor on the prescribed hidden algebra. We establish
a rank obstruction to such factorization and prove its
sufficiency for isometric generators: factorization holds
exactly when the induced hidden-memory channel has Kraus
rank at most the hidden Hilbert-space dimension. The Kraus
rank also gives the minimal intermediate Hilbert-space
dimension. When the hidden and output dimensions are at
least two, the factorizable isometric generators form a
closed subset of measure zero.

For any fixed projective readout, a Mealy generator induces
an operational hidden quantum Markov model in the sense of
Monr\`as, Beige and Wiesner; conversely,
every finite-dimensional operational model admits a Mealy
lift on the same memory space with a rank-one readout.
For their cluster-state instrument, two isometric,
Moore-factorizable lifts with the same instrument and
initial memory state yield identical measured output
statistics but different quantum output states: one has
marginals entangled across $1|2\cdots n$ for $n\geq3$,
with three-output negativity $\tfrac12$, whereas the
other has fully separable marginals. Thus operational
equivalence does not determine output entanglement.


\end{abstract}

\tableofcontents
\clearpage

\section{Introduction}
\label{sec:introduction}

Hidden Markov models are among the basic mathematical frameworks for
systems whose internal dynamics is not directly observable. Since Baum
and Petrie \cite{BaumPetrie1966} they have played a central role in
probability, statistical inference, signal processing, speech
recognition, computational biology and machine learning
\cite{Rabiner1989,EphraimMerhav2002}, and have motivated substantial
developments in filtering, prediction, stochastic realization and
identifiability.

A classical hidden Markov model admits two standard descriptions. In the
\emph{Moore} formulation the output is produced from the hidden state,
separately from the hidden transition; in the \emph{Mealy} formulation
the output is emitted jointly with the transition from the present
hidden state to the next. For finite state and output spaces the two
have the same expressive power at the level of the observable process,
but their realizations need not coincide: passing from one to the other
may require an enlargement of the hidden state space, and minimal
realizations may have different dimensions
\cite{VanluytenWillemsDeMoor2008}. The distinction concerns the
organization of the hidden--output mechanism rather than the observable
process.

The relation between an internal realization and the statistics it
generates is the subject of classical realization theory
\cite{CarlylePaz1971,Anderson1999,Vidyasagar2011Survey,Vidyasagar2014Book}
and of the theory of observable operator models and predictive state
representations
\cite{Jaeger2000,LittmanSuttonSingh2001,ThonJaeger2015}, where a family
of output-labelled operators plays the role of the model and the central
questions are which families generate the same process and how a minimal
realization may be recovered. The present paper is concerned with the
noncommutative counterpart of that passage.

It is useful to recall at the outset a feature of the classical model
which will organize the whole discussion. A Mealy kernel
$p(j,m\mid i)$ makes the hidden variable $X_n$ play two roles at once.
It is the \emph{memory} conditioning the future, appearing as the index
$i$; and it is a \emph{random variable} whose joint law with the emitted
symbols is part of the model, so that quantities such as
$\mathbb P(X_0=i,X_1=j)$ are well defined. Classically the two roles are
filled by one object, because classical information may be copied. We shall see that the Mealy construction retains both roles for the commutative generator associated with a classical Mealy kernel (Theorem~\ref{thm:faithfulness}).

Throughout this paper, \emph{operational HQMM} denotes the
instrument-based hidden quantum Markov model of Monr\`as, Beige and
Wiesner \cite{MonrasBeigeWiesner2011}. We use \emph{Mealy hidden quantum
Markov process} (Mealy HQMP) for our operator-algebraic construction,
specified by a boundary state and a tripartite transition expectation
$\mathcal F$. The word operational
qualifies the instrument description; it does not mean that the
additional output correlations of a Mealy HQMP are unobservable.

In the quantum setting the Moore--Mealy distinction does not transfer
automatically. Transition and emission probabilities are replaced by
completely positive maps on noncommutative algebras, so the
factorization and the order of the emission and hidden-transition
operations become part of the data. Two architectures may generate
identical measured statistics while differing in their internal
correlations, and two orders of the same operations may produce
different joint processes. One must therefore distinguish operational
equivalence, equality of the associated quantum processes, and equality
of their underlying transition mechanisms.

Two approaches are directly relevant. On the operational side, the
operational hidden quantum Markov models of Monr\`as, Beige and Wiesner
\cite{MonrasBeigeWiesner2011} describe the evolution of a hidden quantum
memory by a family of completely positive operations indexed by a
classical alphabet, each determining both the probability of a symbol
and the conditional update of the memory; this is Mealy-type at the
operational level, output and update being carried by the same
output-labelled map. The class has since been studied from several
angles: its correlations have been compared with those of classical
models with the same memory \cite{ONeillBarlowSafranekBeige2012,
AlRasbiClarkBeige2024}, and algorithms have been proposed for fitting
such models to sequence data
\cite{SrinivasanGordonBoots2018,AdhikarySrinivasanGordonBoots2020}. On
the operator-algebraic side, our previous work \cite{AGLS24Q} introduced
hidden quantum Markov processes in which hidden evolution and observable
emission are governed by \emph{distinct} transition expectations, a
Moore-type construction in the terminology adopted here. That framework
extends the theory of algebraic hidden processes
\cite{AccardiSoueidiLuSouissi2024AHP} and is rooted in the
operator-algebraic theory of quantum Markov chains
\cite{accardi1974,AccardiFrigerioLewis1982,ASS20}. Recent work shows
that exchanging the order of hidden propagation and emission may alter
the resulting process \cite{SouissiBarhoumi2026}, confirming that the
internal architecture cannot be ignored.

A Mealy formulation of quantum hidden Markov models based on transition
operation matrices and vector states was developed by Cholewa et al.~\cite{CholewaGawronGlombKurzyk2017}. Their model associates quantum
operations with transitions between classical state labels and focuses
in particular on forward and Viterbi inference. Our construction instead
specifies a joint state on a quasi-local algebra with physical quantum
hidden and output subsystems. The Moore--Mealy question addressed here
concerns factorization of a specified tripartite transition expectation
on the prescribed hidden algebra.

Quantum Moore and Mealy machines have also been studied as transducers.
Tamon and Xie \cite{TamonXie2019} established an operational equivalence
after a suitable enlargement of the state space; their result concerns
input--output equivalence and equality of output probabilities.
Their construction appends a classical output register through
$\rho\mapsto\sum_b|b\rangle\langle b|\otimes\Lambda^{\mathrm{TX}}_{a,b}(\rho)$,
whose target space has dimension $|O|d_H$. This is operationally
analogous to recording the emitted symbol, but is not an identification
with the coherent site-storing lift below. In particular, an enlarged
operational realization does not contradict our obstruction to
factorization on a fixed hidden algebra.
The question addressed here is different. We work with states on a
quasi-local hidden--observable algebra, and we ask not only which
architectures produce the same output statistics, but what is retained
and what is lost in passing from the full state to those statistics.

The two levels of description are mathematically distinct objects. An
operational hidden quantum Markov model determines the statistics of a
measured output process together with the conditional evolution of a
single memory. An operator-algebraic hidden process determines a state
on a larger noncommutative algebra, and therefore retains expectation
values of joint observables involving hidden and output systems at
several times. The purpose of this paper is to place both within a common framework, to relate the two levels of description, and to show which hidden--output correlations the operator-algebraic process specifies beyond the operational instrument.

We emphasize at once what is \emph{not} claimed. At the level of
classical output distributions the two formalisms have exactly the same
expressive power, since every operational hidden quantum Markov model is the
observable reduction of a Mealy process on the same hidden space
(Theorem~\ref{thm:mealy-lift}). Nor do we address the comparison between
quantum and classical memory requirements for generating a given
classical process, a separate line of work in which the relevant
quantities are the quantum memory dimension and the classical
statistical complexity \cite{CrutchfieldYoung1989,
GuWiesnerRieperVedral2012,BinderThompsonGu2018,ZonniosBoydBinder2025}.

The construction belongs to the theory of finitely correlated
states \cite{FannesNachtergaeleWerner1992}. Write $\mathcal A_H=\mathcal B(\mathcal H_H)$ and
$\mathcal A_O=\mathcal B(\mathcal H_O)$, with
$d_H=\dim\mathcal H_H$ and $d_O=\dim\mathcal H_O$. The transition expectation
\[
\mathcal F:
(\mathcal A_H\otimes\mathcal A_O)\otimes\mathcal A_H
\longrightarrow\mathcal A_H
\]
has physical algebra $\mathcal A_H\otimes\mathcal A_O$ and
auxiliary algebra $\mathcal A_H$. We use its one-sided
boundary-state form, without assuming translation invariance.

The extra structure here is the distinguished decomposition of each
physical site into hidden and output factors, with the auxiliary
memory algebra identified with the prescribed hidden algebra. This
poses a constrained realization problem: whether a given tripartite
transition expectation admits a Moore factorization on that same
hidden algebra. The question concerns the specified generator, not
all possible realizations of the resulting process.

For an isometric generator
$W:\mathcal H_H\to\mathcal H_H\otimes\mathcal H_O\otimes\mathcal H_H$,
as used in the sequential generation of matrix product states
\cite{SchonEtAl2005,PerezGarciaEtAl2007}, we characterize such
factorization in terms of the dimension of the span of the matrix
blocks of $W$, equivalently the Schmidt rank of $W$ as an element of
$(\mathcal H_H\otimes\mathcal H_O)\otimes\mathcal B(\mathcal H_H)$
\cite{NielsenEtAl2003}.
We also identify the minimal intermediate Hilbert-space dimension
among completely positive factorizations. For a fixed projective
readout, observable reduction maps each Mealy generator to a quantum
instrument; that instrument need not determine the entanglement of
the emitted output systems.

Thus, our contribution is not a new class of finitely correlated
states, but structural results for generators with a prescribed
hidden--output decomposition, including the characterization of
isometric Moore factorization and results on the information
preserved by, or lost under, operational reduction at a fixed readout.
\subsection*{Results}

\emph{Moore factorization.}
Our main structural result characterizes Moore factorization
for isometric Mealy generators. Let
\[
W:\mathcal H_H\longrightarrow
\mathcal H_H\otimes\mathcal H_O\otimes\mathcal H_H,
\qquad
\mathcal F=\operatorname{Ad}_W,
\]
and write
\[
W\psi=\sum_{i,m}|i\rangle\otimes|m\rangle\otimes W_{i,m}\psi.
\]
Then $\mathcal F$ admits a factorization
\[
\mathcal F=E_H\circ
(E_O\otimes\operatorname{id}_{\mathcal A_H})
\]
by completely positive unital maps on the prescribed hidden
algebra if and only if
\[
\dim\operatorname{span}\{W_{i,m}\}\leq d_H
\]
(Theorem~\ref{thm:isometric-moore-characterization}).
Whenever a factorization exists, the factors may be chosen
isometric. More generally, allowing the intermediate
Hilbert space to vary, its minimal dimension is exactly
$\dim\operatorname{span}\{W_{i,m}\}$
(Corollary~\ref{cor:minimal-intermediate-dimension}). This dimension
is the Kraus rank of the hidden channel
$\mathcal E(x)=\mathcal F(I_H\otimes I_O\otimes x)$, so the criterion
depends on $\mathcal F$ only through $\mathcal E$
(Corollary~\ref{cor:factorization-channel}).

The necessity follows from a rank invariant
$r_\star(\mathcal F)$, which satisfies
$r_\star(\mathcal F)\leq d_H^2$ for every Moore-factorizable
generator and equals
$(\dim\operatorname{span}\{W_{i,m}\})^2$ in the isometric case
(Propositions~\ref{prop:rank-obstruction}
and~\ref{prop:rstar-block}). Sufficiency follows from
a Schmidt decomposition of $W$.
Explicit nonfactorizable generators occur already for
$d_H=d_O=2$
(Example~\ref{ex:equal-dimension-non-moore}), and for $d_H,d_O\geq2$
the factorizable isometric generators form a closed set of measure
zero (Theorem~\ref{thm:moore-nongeneric}).
The factorizing generators recover the Moore construction
of \cite{AGLS24Q}
(Theorem~\ref{thm:moore-mealy-identification}).
These assertions concern the specified generator:
failure of its factorization does not exclude another
Moore-factorizable generator of the same process.

\emph{The associated process.}
A Mealy transition expectation and a boundary state determine
a unique quasi-local hidden--output state
(Theorem~\ref{thm:mealy-existence}).
We characterize shift invariance by agreement of the boundary
state and its image under the hidden channel on the range
of the iterated transition maps
(Proposition~\ref{prop:stationarity}).
An invariant boundary state always exists and is sufficient
for stationarity, but need not be necessary
(Example~\ref{ex:stationarity-counterexample}).
The hidden marginal is a quantum Markov chain
(Theorem~\ref{thm:hidden-marginal}), whereas the observable
marginal is a finitely correlated state with MPDO bond
dimension at most $d_H^2$
(Theorem~\ref{thm:observable-marginal-fcs}).

\emph{The observable reduction.} Fixing a projective readout
$\{b_m\}_{m\in O}$, we associate with $\mathcal F$ the maps
$\Phi_m=\mathcal F(I_H\otimes b_m\otimes\,\cdot\,)$, prove that their
Schr\"odinger duals form a quantum instrument, and show that the
resulting word probabilities reproduce exactly the observable cylinder
probabilities of the process
(Theorem~\ref{thm:observable-reduction}). The reduction is surjective,
by completely positive lifts (Theorem~\ref{thm:mealy-lift}); isometric
lifts exist exactly when each operation has at most $d_H$ Kraus
operators (Remark~\ref{rem:lift-no-obstruction}), and whether they
admit a Moore factorization depends only on the instrument, through its
hidden channel (Corollary~\ref{cor:factorization-channel}). The
reduction is defined on
generators, not on process states: different generators of the same
state can have different reductions
(Example~\ref{ex:stationarity-counterexample}). It is not
injective: an erasing and a site-storing lift of the same instrument
have hidden site marginals of entropy $0$ and $H(p)$ respectively, where
$H(p)$ is the Shannon entropy of the first output symbol
(Corollary~\ref{cor:channel-not-chain}). What the reduction does
determine is the hidden channel, hence whether the boundary state is
invariant for it.

\emph{Faithfulness to the classical model.}
For the commutative generator associated with a classical
Mealy kernel, the construction recovers the full joint
hidden--output law (Theorem~\ref{thm:faithfulness}).
A commutative instrument also determines this law through
its transition kernel and initial distribution.
For a quantum memory, the instrument alone does not specify
an unmeasured tensor-product state of physical hidden sites.

\emph{The cluster state.} The central example of
\cite{MonrasBeigeWiesner2011} shows what the operational formalism does
not record. The instrument describing the non-adaptive read-out of a
one-dimensional cluster state lifts to a Mealy process whose emitted
systems carry the marginals of the cluster state, entangled from three
emitted systems on, with negativity $\tfrac12$ across the cut
$\{1\}|\{2,3\}$, and to a Mealy process whose emitted marginals are
diagonal in a product basis (Theorem~\ref{thm:cluster-separation}). The
entanglement of the emitted systems is therefore not a function of the
operational hidden quantum Markov model.

\subsection*{Organization}
Section~\ref{sec:mealy-hqmp} constructs the quasi-local process,
analyses shift invariance, and identifies its hidden and
observable marginals. Section~\ref{sec:moore-as-mealy}
identifies Moore processes with processes admitting factorizing
generators, establishes a rank obstruction, and proves its
sufficiency for isometric generators.
Section~\ref{sec:observable-reduction} develops the observable reduction
and proves the lifting theorem. Section~\ref{sec:faithfulness} establishes
faithfulness to the classical model. Section~\ref{sec:cluster} treats
the cluster-state example of \cite{MonrasBeigeWiesner2011}.
Section~\ref{sec:conclusion} concludes and lists open problems.

\section{Preliminaries}
\label{sec:preliminaries}

All Hilbert spaces and local algebras are finite dimensional.
Inner products are conjugate-linear in the first variable, and
all tensor products of local algebras are spatial tensor products.
We use the standard facts on completely positive maps, quantum
Markov chains and finitely correlated states collected in
\cite{Choi1975,Kraus1983,AccardiFrigerioLewis1982,
FannesNachtergaeleWerner1992,ASS20,
AccardiSoueidiLuSouissi2024AHP,AGLS24Q}.

\subsection{States and completely positive maps}
\label{subsec:states}
\label{subsec:cp-maps}

For a Hilbert space $\mathcal H$, write $\mathcal B(\mathcal H)$
for its operator algebra, $I_{\mathcal H}$ for the identity, and
$\mathcal T(\mathcal H)$ for the same vector space viewed as
trace-class operators. Every state has the form
$\omega(x)=\operatorname{Tr}(\rho_\omega x)$ with a density operator
$\rho_\omega$, and we identify states with their density operators
when convenient; restrictions to tensor factors correspond to partial
traces. The von Neumann entropy is
$S(\rho)=-\operatorname{Tr}(\rho\log_2\rho)$, with $0\log_2 0=0$.
Throughout the quantum construction, put
\[
\mathcal A_H=\mathcal B(\mathcal H_H),\qquad
\mathcal A_O=\mathcal B(\mathcal H_O),\qquad
d_H=\dim\mathcal H_H,\quad d_O=\dim\mathcal H_O,
\]
and denote the finite output alphabet by $O$.
A complete rank-one readout has $|O|=d_O$; the additional
condition $|O|\leq d_H$ is needed only when orthonormal
hidden-site records of all output symbols are required.

For a linear map
$\Phi:\mathcal B(\mathcal H_{\mathrm{out}})
\to\mathcal B(\mathcal H_{\mathrm{in}})$, its predual is
characterized by
\begin{equation}
\label{eq:predual}
\operatorname{Tr}\bigl(\Phi_*(\rho)x\bigr)
=\operatorname{Tr}\bigl(\rho\Phi(x)\bigr).
\end{equation}
For a map $\Psi$ on states, its dual is denoted by $\Psi^*$;
thus $(\Phi_*)^*=\Phi$ and $(\Psi^*)_*=\Psi$. The map $\Phi$ is
unital if $\Phi(I)=I$ and subunital if $\Phi(I)\leq I$. By the
Choi--Kraus theorem \cite{Choi1975,Kraus1983}, $\Phi$ is completely
positive (CP) if and only if
\begin{equation}
\label{eq:choi-kraus}
\Phi(x)=\sum_\alpha V_\alpha^*xV_\alpha,
\qquad
V_\alpha:\mathcal H_{\mathrm{in}}
\longrightarrow\mathcal H_{\mathrm{out}},
\end{equation}
for a finite family $(V_\alpha)_\alpha$. Then
$\Phi_*(\rho)=\sum_\alpha V_\alpha\rho V_\alpha^*$, and $\Phi$ is
unital, respectively subunital, if and only if
$\sum_\alpha V_\alpha^*V_\alpha=I$, respectively
$\sum_\alpha V_\alpha^*V_\alpha\leq I$.
A CP trace-preserving map on states is a \emph{quantum channel};
a CP map that is trace non-increasing on positive operators
is a \emph{quantum operation}. By duality, these correspond
to CP unital and CP subunital maps, respectively. Every quantum
channel on $\mathcal T(\mathcal H)$ has an invariant density
operator, by Brouwer's fixed point theorem applied to the compact
convex set of density operators.

For an isometry $W:\mathcal H\to\mathcal K$, the map
$\operatorname{Ad}_W(x)=W^*xW$ is CP and unital, with predual
$\rho\mapsto W\rho W^*$, and
$\operatorname{Ad}_W\circ\operatorname{Ad}_{W'}
=\operatorname{Ad}_{W'W}$ for an isometry
$W':\mathcal K\to\mathcal L$. We call a CP unital map
\emph{isometric} if it is of the form $\operatorname{Ad}_W$; this
terminology refers to its implementing operator $W$, not to
preservation of operator norms.

\subsection{Quantum Markov chains and finitely correlated states}
\label{subsec:quasilocal}
\label{subsec:qmc}

If finite-dimensional unital $C^*$-algebras $(\mathcal C_n)_{n\geq0}$
are identified through unital injective $*$-homomorphisms
$\iota_n:\mathcal C_n\to\mathcal C_{n+1}$, their quasi-local algebra
is $\mathcal C_\infty=\overline{\bigcup_n\mathcal C_n}^{\|\cdot\|}$,
and states $\phi_n$ on $\mathcal C_n$ satisfying
$\phi_{n+1}\circ\iota_n=\phi_n$ extend uniquely, by continuity, to a
state on $\mathcal C_\infty$.

\begin{definition}
\label{def:transition-expectation}
\label{def:qmc}
A \emph{transition expectation} on a finite-dimensional
unital $C^*$-algebra $\mathcal A$ is a CP unital map
$E:\mathcal A\otimes\mathcal A\to\mathcal A$. On
$\mathcal A^{\otimes(n+1)}$, put
\begin{equation}
\label{eq:qmc-iteration}
E_{[0,0]}(a)=E(a\otimes I),\qquad
E_{[0,n]}=E\circ(\operatorname{id}_{\mathcal A}
\otimes E_{[0,n-1]}),\quad n\geq1.
\end{equation}
For a state $\omega$ on $\mathcal A$, the states
$\omega\circ E_{[0,n]}$ are compatible under $x\mapsto x\otimes I$,
since $E(I\otimes I)=I$. The state they define on
$\overline{\bigcup_n\mathcal A^{\otimes(n+1)}}$ is the
\emph{quantum Markov chain} generated by $(\omega,E)$
\cite{accardi1974,AccardiFrigerioLewis1982,ASS20}. The terminal map
$a\mapsto E(a\otimes I)$ is not assumed to be the identity.
\end{definition}

\begin{remark}
\label{rem:fcs-general}
The same construction applies to a CP unital map
$E:\mathcal C\otimes\mathcal B\to\mathcal B$, with
finite-dimensional unital $C^*$-algebras $\mathcal C,
\mathcal B$ and a boundary state $\omega$ on $\mathcal B$.
For $E_c(x)=E(c\otimes x)$, its local expectations are
\[
\phi(c_0\otimes\cdots\otimes c_n)
=\omega\bigl(E_{c_0}\circ\cdots\circ E_{c_n}(I_{\mathcal B})\bigr).
\]
We call this the \emph{finitely correlated state} generated
by $(\omega,E)$, with physical algebra $\mathcal C$ and
auxiliary algebra $\mathcal B$. This is the one-sided
boundary-state form of the construction in
\cite{FannesNachtergaeleWerner1992}; no invariance of $\omega$ is
assumed. By MPDO bond dimension we mean the virtual dimension in
an open-boundary matrix-product operator representation
of a density operator, rather than the Hilbert-space
dimension of the auxiliary memory.
\end{remark}

\subsection{Classical Mealy hidden Markov models}
\label{subsec:classical-mealy}

We use the finite-state Mealy convention of
\cite{VanluytenWillemsDeMoor2008,Vidyasagar2011Survey,
Vidyasagar2014Book}. Let $S$ be a finite hidden state space
and $O$ a finite output alphabet. Probability distributions
on $S$ are row vectors.

\begin{definition}
\label{def:classical-mealy}
A classical Mealy hidden Markov model consists of an
initial distribution $\pi$ on $S$ and a kernel
$p(j,m\mid i)\geq0$, the conditional probability of moving to
hidden state $j$ and emitting $m$ given the current hidden state
$i$, satisfying
\begin{equation}
\label{eq:mealy-normalization}
\sum_{m\in O}\sum_{j\in S}p(j,m\mid i)=1,
\qquad i\in S.
\end{equation}
\end{definition}

The joint hidden--output law is
\begin{equation}
\label{eq:classical-trajectory-law}
\mathbb P(X_0=i_0,Y_1=m_1,X_1=i_1,\ldots,Y_n=m_n,X_n=i_n)
=\pi_{i_0}\prod_{k=1}^n p(i_k,m_k\mid i_{k-1}).
\end{equation}
The output-labelled matrices
\begin{equation}
\label{eq:output-labelled-matrix}
P_m=(p(j,m\mid i))_{i,j\in S}
\end{equation}
are substochastic, and $P=\sum_mP_m$ is row-stochastic:
$P\mathbf1=\mathbf1$. Summing
\eqref{eq:classical-trajectory-law} over the hidden states gives the
word probabilities
\begin{equation}
\label{eq:classical-word-probability}
\mathbb P(m_1\cdots m_n)=\pi P_{m_1}\cdots P_{m_n}\mathbf1,
\end{equation}
which are consistent, $\sum_{m\in O}\mathbb P(wm)=\mathbb P(w)$,
because $P\mathbf1=\mathbf1$.

\subsection{Operational hidden quantum Markov models}
\label{subsec:hqmm}

The operational model of Monr\`as, Beige and Wiesner
\cite{MonrasBeigeWiesner2011} replaces the output-labelled
substochastic matrices by quantum operations on a memory.

\begin{definition}
\label{def:hqmm}
An \emph{operational hidden quantum Markov model} over $O$
is a triple $(\mathcal H,\rho_0,\{\mathcal K_m\}_{m\in O})$,
where $\rho_0$ is a density operator on $\mathcal H$,
each $\mathcal K_m:\mathcal T(\mathcal H)
\to\mathcal T(\mathcal H)$ is CP and trace non-increasing,
and $\sum_m\mathcal K_m$ is trace preserving.
Thus $\{\mathcal K_m\}$ is a finite-outcome quantum instrument.
\end{definition}

Choose Kraus representations
\begin{equation}
\label{eq:hqmm-kraus}
\mathcal K_m(\rho)=\sum_\alpha
L_{m,\alpha}\rho L_{m,\alpha}^*.
\end{equation}
The instrument normalization is equivalent to
\begin{equation}
\label{eq:hqmm-kraus-completeness}
\sum_{m,\alpha}L_{m,\alpha}^*L_{m,\alpha}=I_{\mathcal H}.
\end{equation}
Writing $\Phi_m=(\mathcal K_m)^*$, one has
$\Phi_m(x)=\sum_\alpha L_{m,\alpha}^*xL_{m,\alpha}$ and
$\sum_m\Phi_m(I)=I$.
For a word $w=m_1\cdots m_n$, duality gives
\[
\mathbb P_{\rho_0}(w)
=\operatorname{Tr}\bigl[
\mathcal K_{m_n}\circ\cdots\circ\mathcal K_{m_1}(\rho_0)
\bigr]
=\operatorname{Tr}\bigl[
\rho_0\,\Phi_{m_1}\circ\cdots\circ\Phi_{m_n}(I)
\bigr].
\]
When this probability is positive, the conditional memory
state is
$\mathcal K_{m_n}\circ\cdots\circ\mathcal K_{m_1}(\rho_0)
/\mathbb P_{\rho_0}(w)$.

\section{Mealy Hidden Quantum Markov Processes}
\label{sec:mealy-hqmp}

In the Moore construction of \cite{AGLS24Q}, the hidden transition and
the emission are encoded by separate transition expectations. In a
classical Mealy model, the passage from the present hidden state to the
next one and the emission of the observable symbol form a single
probabilistic event. This section formulates that one-step mechanism in
operator-algebraic terms, with no factorization between hidden
propagation and emission imposed by the definition.

\subsection{Operator-algebraic formulation of the classical Mealy kernel}
\label{subsec:classical-mealy-te}

Let $p(j,m\mid i)$ be the kernel of a classical Mealy model
(Section~\ref{subsec:classical-mealy}), and let $C(S)$ and $C(O)$ be
the commutative $C^{*}$-algebras of complex functions on $S$ and $O$.
Using the canonical identification
$C(S)\otimes C(O)\otimes C(S)\simeq C(S\times O\times S)$, define
\begin{equation}
\label{eq:classical-mealy-te}
\mathcal F_{\mathrm{cl}}:
C(S)\otimes C(O)\otimes C(S)\longrightarrow C(S),
\qquad
\bigl(\mathcal F_{\mathrm{cl}}\eta\bigr)(i)
=\sum_{m\in O}\sum_{j\in S}p(j,m\mid i)\,\eta(i,m,j).
\end{equation}

\begin{proposition}
\label{prop:classical-mealy-te}
The map $\mathcal F_{\mathrm{cl}}$ is completely positive and unital, and it
determines the classical Mealy kernel uniquely.
\end{proposition}

\begin{proof}
The map is positive, hence completely positive since its domain is
commutative, and it is unital by \eqref{eq:mealy-normalization}. If
$e_i\in C(S)$ and $\chi_m\in C(O)$ denote the indicator functions, then
\begin{equation}
\label{eq:classical-kernel-recovery}
\mathcal F_{\mathrm{cl}}\bigl(e_i\otimes\chi_m\otimes e_j\bigr)
=p(j,m\mid i)\,e_i ,
\end{equation}
so every coefficient is recovered from $\mathcal F_{\mathrm{cl}}$.
\end{proof}

The three tensor factors represent the present hidden state, the
emitted output and the next hidden state, so the one-step mechanism of
a classical Mealy model is a single transition expectation.

\subsection{Mealy transition expectations}
\label{subsec:mealy-te}

\begin{definition}
\label{def:mealy-te}
A \emph{Mealy transition expectation} on
$(\mathcal A_H,\mathcal A_O)$ is a completely positive unital map
\[
\mathcal F:\mathcal A_H\otimes\mathcal A_O\otimes\mathcal A_H
\longrightarrow\mathcal A_H .
\]
\end{definition}

The map $\mathcal F$ is understood in the Heisenberg picture: it sends a joint
observable of one hidden--output--hidden step to an observable of the
hidden boundary algebra. By \eqref{eq:predual}, its predual
\begin{equation}
\label{eq:mealy-te-predual}
\mathcal F_*:\mathcal T(\mathcal H_H)
\longrightarrow
\mathcal T\bigl(\mathcal H_H\otimes\mathcal H_O\otimes\mathcal H_H\bigr)
\end{equation}
is a quantum channel. Unlike the Moore construction, a Mealy transition
expectation is not required to decompose into separate hidden-transition
and emission maps; whether such a factorization exists is a structural
property of $\mathcal F$, studied in Section~\ref{sec:moore-as-mealy}.

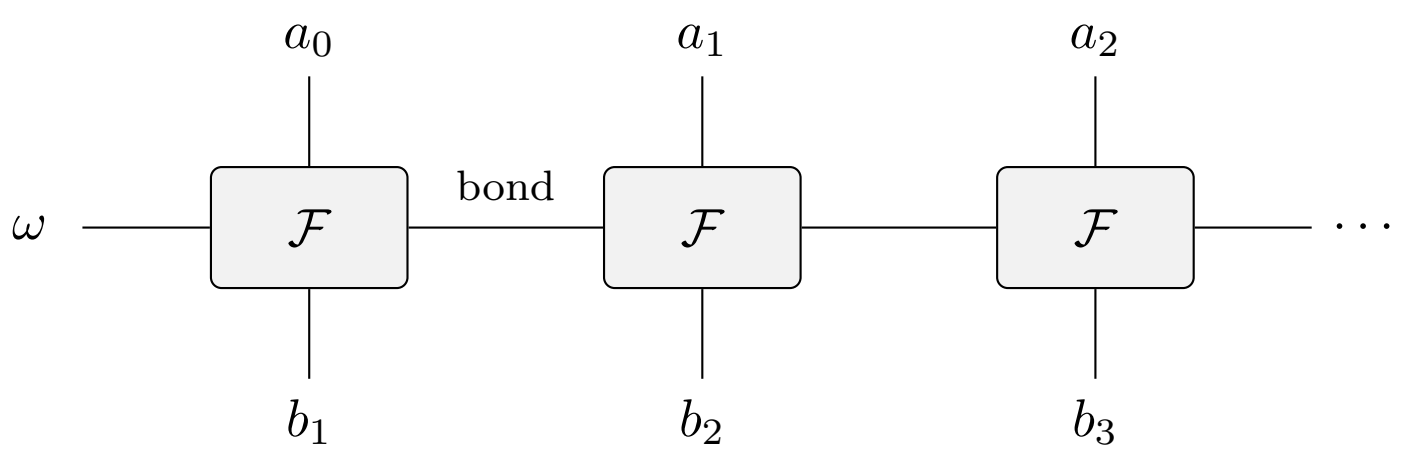
\begin{figure}[t]
\centering
\begin{tikzpicture}[
  every node/.style={font=\small},
  box/.style={draw, rounded corners=2pt, minimum width=13mm,
              minimum height=8mm, fill=black!5},
  line/.style={-}
]
\node[box] (T0) at (0,0) {$\mathcal F$};
\node[box] (T1) at (2.6,0) {$\mathcal F$};
\node[box] (T2) at (5.2,0) {$\mathcal F$};
\node        (dots) at (7.0,0) {$\cdots$};

\draw[line] (-1.5,0) -- (T0);
\node at (-1.85,0) {$\omega$};
\draw[line] (T0) -- (T1);
\draw[line] (T1) -- (T2);
\draw[line] (T2) -- (dots);

\draw[line] (T0) -- ++(0,1.0) node[above] {$a_0$};
\draw[line] (T1) -- ++(0,1.0) node[above] {$a_1$};
\draw[line] (T2) -- ++(0,1.0) node[above] {$a_2$};

\draw[line] (T0) -- ++(0,-1.0) node[below] {$b_1$};
\draw[line] (T1) -- ++(0,-1.0) node[below] {$b_2$};
\draw[line] (T2) -- ++(0,-1.0) node[below] {$b_3$};

\node at (1.3,0.28) {\scriptsize bond};
\end{tikzpicture}
\caption{One Mealy transition expectation and its iteration
\eqref{eq:mealy-iteration-tensors}. The horizontal line is the bond, the
upward legs carry the hidden site observables $a_k$, the downward legs
the emitted output observables $b_{k+1}$. The boundary state $\omega$
closes the chain on the left.}
\label{fig:mealy-step}
\end{figure}

\begin{remark}
\label{rem:bond-site}
Under the iteration of Section~\ref{subsec:local-algebras}, the three
factors of $\mathcal F$ play different roles. The first carries the
hidden observable of the current site and the second the output
observable of the current step. The third receives the value of the
whole tail of the recursion \eqref{eq:mealy-iteration}, a single
element of $\mathcal A_H$; we call it the \emph{bond}, and no site
algebra is ever evaluated in it. In the Schr\"odinger picture,
\eqref{eq:mealy-te-predual} maps a state of the bond to a state of
$\mathcal H_H\otimes\mathcal H_O\otimes\mathcal H_H$: it emits one
hidden site and one output and returns a new bond, which is not a
subsystem of the quasi-local algebra constructed below
(Figure~\ref{fig:mealy-step}). The observable reduction of
Section~\ref{sec:observable-reduction} evaluates the bond argument with
the identity in the site argument, while the boundary map
$\mathcal F_\partial$ of \eqref{eq:mealy-T0} propagates a site
observable to the bond, which is why it appears in every correlation
function of the hidden sites
(Proposition~\ref{prop:hidden-correlations}).
\end{remark}

\subsection{The quasi-local process}
\label{subsec:local-algebras}
\label{subsec:finite-volume}

Let $\mathcal A_n^H$ ($n\geq0$) and $\mathcal A_n^O$ ($n\geq1$)
be fixed copies of $\mathcal A_H$ and $\mathcal A_O$, respectively,
and put
\begin{equation}
\label{eq:local-algebra}
\mathcal A_{[0,n]}
=\mathcal A_0^H\otimes
\bigotimes_{k=1}^{n}
\bigl(\mathcal A_k^O\otimes\mathcal A_k^H\bigr),
\qquad n\geq0.
\end{equation}
These algebras increase through the unital $*$-monomorphisms
\begin{equation}
\label{eq:mealy-embedding}
\iota_n:\mathcal A_{[0,n]}\longrightarrow\mathcal A_{[0,n+1]},
\qquad
\iota_n(X)=X\otimes I_O\otimes I_H,
\end{equation}
with inductive limit
\begin{equation}
\label{eq:mealy-quasilocal}
\mathcal A_\infty
=\overline{\bigcup_{n\geq0}\mathcal A_{[0,n]}}^{\,\|\cdot\|}.
\end{equation}
Maps on local algebras are transported to shifted blocks through the
fixed identifications of the corresponding tensor factors.

Let $\mathcal F$ be a Mealy transition expectation, and define its
\emph{boundary map}
\begin{equation}
\label{eq:mealy-T0}
\mathcal F_\partial:\mathcal A_H\longrightarrow\mathcal A_H,
\qquad
\mathcal F_\partial(a)
=\mathcal F(a\otimes I_O\otimes I_H).
\end{equation}
Starting with $\mathcal F_{[0,0]}=\mathcal F_\partial$, define
$\mathcal F_{[0,n]}:\mathcal A_{[0,n]}\to\mathcal A_H$
recursively by
\begin{equation}
\label{eq:mealy-iteration}
\mathcal F_{[0,n]}
=\mathcal F\circ
\bigl(
\operatorname{id}_{\mathcal A_H}
\otimes\operatorname{id}_{\mathcal A_O}
\otimes\mathcal F_{[0,n-1]}
\bigr),
\qquad n\geq1,
\end{equation}
where $\mathcal F_{[0,n-1]}$ acts on the factors following
$\mathcal A_0^H\otimes\mathcal A_1^O$. In particular,
$\mathcal F_{[0,1]}(a_0\otimes b_1\otimes a_1)
=\mathcal F\bigl(a_0\otimes b_1\otimes\mathcal F_\partial(a_1)\bigr)$,
and on elementary tensors
\begin{equation}
\label{eq:mealy-iteration-tensors}
\begin{aligned}
&\mathcal F_{[0,n]}
\bigl(a_0\otimes b_1\otimes a_1\otimes\cdots
      \otimes b_n\otimes a_n\bigr)
\\
&\quad=
\mathcal F\bigl(
a_0\otimes b_1\otimes
\mathcal F\bigl(
a_1\otimes b_2\otimes\cdots
\mathcal F\bigl(
a_{n-1}\otimes b_n\otimes\mathcal F_\partial(a_n)
\bigr)\cdots\bigr)\bigr),
\qquad n\geq2.
\end{aligned}
\end{equation}

\begin{theorem}
\label{thm:mealy-existence}
Let $\omega$ be a state on $\mathcal A_H$. For every $n\geq0$, the map
$\mathcal F_{[0,n]}$ is completely positive and unital, and
\begin{equation}
\label{eq:compatibility}
\mathcal F_{[0,n+1]}(X\otimes I_O\otimes I_H)=\mathcal F_{[0,n]}(X),
\qquad X\in\mathcal A_{[0,n]}.
\end{equation}
Consequently, there is a unique state $\varphi$ on $\mathcal A_\infty$
such that
\begin{equation}
\label{eq:finite-volume-state}
\varphi|_{\mathcal A_{[0,n]}}
=\omega\circ\mathcal F_{[0,n]},
\qquad n\geq0.
\end{equation}
\end{theorem}

\begin{proof}
Complete positivity and unitality follow by induction from
\eqref{eq:mealy-iteration}, since $a\mapsto a\otimes I_O\otimes I_H$ is
a unital $*$-homomorphism and both properties are stable under tensor
products and composition. Unitality of $\mathcal F$ gives
\begin{equation}
\label{eq:T0-unital}
\mathcal F_\partial(I_H)=\mathcal F(I_H\otimes I_O\otimes I_H)=I_H ,
\end{equation}
hence
$\mathcal F_{[0,1]}(a\otimes I_O\otimes I_H)
=\mathcal F(a\otimes I_O\otimes I_H)=\mathcal F_{[0,0]}(a)$.
If \eqref{eq:compatibility} holds for $n-1$ and
$X=a_0\otimes b_1\otimes Y$, then
\[
\begin{aligned}
\mathcal F_{[0,n+1]}(X\otimes I_O\otimes I_H)
&=\mathcal F\bigl(a_0\otimes b_1\otimes
\mathcal F_{[0,n]}(Y\otimes I_O\otimes I_H)\bigr)\\
&=\mathcal F\bigl(a_0\otimes b_1\otimes
\mathcal F_{[0,n-1]}(Y)\bigr)
=\mathcal F_{[0,n]}(X),
\end{aligned}
\]
and linearity gives \eqref{eq:compatibility}. The states
$\omega\circ\mathcal F_{[0,n]}$ are therefore compatible with
\eqref{eq:mealy-embedding}, and they extend uniquely to
$\mathcal A_\infty$ (Section~\ref{subsec:quasilocal}).
\end{proof}

\begin{definition}
\label{def:mealy-process}
The state $\varphi$ of Theorem~\ref{thm:mealy-existence} is the
\emph{Mealy hidden quantum Markov process} associated with
$(\omega,\mathcal F)$, and $\omega$ is its \emph{boundary state}.
\end{definition}

\subsection{The hidden channel and stationarity}
\label{subsec:slot-maps}

\begin{definition}
\label{def:hidden-slot-maps}
For each $a\in\mathcal A_H$, define
\begin{equation}
\label{eq:slot-maps}
\mathcal E_a:\mathcal A_H\longrightarrow\mathcal A_H,
\qquad
\mathcal E_a(x)=\mathcal F(a\otimes I_O\otimes x).
\end{equation}
We call
\begin{equation}
\label{eq:hidden-channel}
\mathcal E:=\mathcal E_{I_H},
\qquad
\mathcal E(x)=\mathcal F(I_H\otimes I_O\otimes x),
\end{equation}
the \emph{hidden channel} associated with $\mathcal F$, written
in the Heisenberg picture.
\end{definition}

The map $\mathcal E$ is CP and unital, and its predual
$\mathcal E_*(\rho)=\operatorname{Tr}_{1,2}\bigl(\mathcal F_*(\rho)\bigr)$,
where $\operatorname{Tr}_{1,2}$ traces out the hidden site and the
output, is the reduced evolution of the bond. For the classical
generator \eqref{eq:classical-mealy-te},
$(\mathcal E_{\mathrm{cl}}h)(i)=\sum_{j\in S}q(j\mid i)h(j)$ with
$q(j\mid i)=\sum_{m\in O}p(j,m\mid i)$, the transition operator of the
hidden Markov chain. The assignment $a\mapsto\mathcal E_a$ is linear,
$\mathcal E_a(I_H)=\mathcal F_\partial(a)$, and $\mathcal E_a$ is CP
for $a\geq0$. The hidden channel describes the propagation of the bond,
whereas the maps $\mathcal E_a$ also encode the dependence on
hidden-site observables.

\begin{remark}
\label{rem:boundary-state}
The boundary state is not the initial hidden marginal: by
\eqref{eq:finite-volume-state},
$\varphi|_{\mathcal A_0^H}=\omega\circ \mathcal F_\partial$, which
coincides with $\omega$ if and only if
$\omega\circ \mathcal F_\partial=\omega$. This condition differs from
the invariance condition $\omega\circ\mathcal E=\omega$, which is
sufficient for shift invariance by Proposition~\ref{prop:stationarity};
the two involve $a\mapsto\mathcal E_a$ at different points, since
$\mathcal F_\partial(a)=\mathcal E_a(I_H)$ whereas
$\mathcal E=\mathcal E_{I_H}$.
\end{remark}

The maps $\theta_n:\mathcal A_{[0,n]}\to\mathcal A_{[0,n+1]}$,
$\theta_n(X)=I_H\otimes I_O\otimes X$, where each tensor factor of $X$
is placed at the site with index increased by one, are unital injective
$*$-homomorphisms with $\theta_{n+1}\circ\iota_n=\iota_{n+1}\circ\theta_n$.
They therefore define a unital injective $*$-endomorphism $\theta$ of
$\mathcal A_\infty$, the \emph{shift}. A state $\varphi$ is called
\emph{stationary}, or \emph{shift invariant}, if
$\varphi\circ\theta=\varphi$.

\begin{proposition}[Stationarity]
\label{prop:stationarity}
Let $\varphi$ be the Mealy process generated by
$(\omega,\mathcal F)$, let $\mathcal E$ be its hidden channel,
and let $\rho_\omega$ be the density operator of $\omega$.
Set
\begin{equation}
\label{eq:range-space}
\mathcal R_{\mathcal F}
=\bigcup_{n\geq0}
\mathcal F_{[0,n]}(\mathcal A_{[0,n]}).
\end{equation}
Then:
\begin{enumerate}
\item[(a)]
$\omega\circ\mathcal E=\omega$ if and only if
$\mathcal E_*(\rho_\omega)=\rho_\omega$.

\item[(b)]
$\mathcal R_{\mathcal F}$ is a linear subspace of
$\mathcal A_H$ containing $I_H$, and
\[
\varphi\circ\theta=\varphi
\quad\Longleftrightarrow\quad
(\omega\circ\mathcal E)|_{\mathcal R_{\mathcal F}}
=\omega|_{\mathcal R_{\mathcal F}}.
\]

\item[(c)]
The condition $\omega\circ\mathcal E=\omega$ is sufficient
for stationarity. It is also necessary when
$\mathcal R_{\mathcal F}=\mathcal A_H$, in particular when
$\mathcal F_\partial$ is surjective.

\item[(d)]
There exists a state $\omega$ satisfying
$\omega\circ\mathcal E=\omega$.
\end{enumerate}
\end{proposition}

\begin{proof}
Part (a) follows from
\[
(\omega\circ\mathcal E)(x)
=\operatorname{Tr}\bigl(\mathcal E_*(\rho_\omega)x\bigr),
\qquad x\in\mathcal A_H.
\]

For (b), compatibility gives
\[
\mathcal F_{[0,n]}(\mathcal A_{[0,n]})
\subseteq
\mathcal F_{[0,n+1]}(\mathcal A_{[0,n+1]}).
\]
Thus $\mathcal R_{\mathcal F}$ is an increasing union of
linear subspaces, each containing $I_H$. For
$X\in\mathcal A_{[0,n]}$, the recursion yields
\[
\mathcal F_{[0,n+1]}(\theta_n(X))
=\mathcal E\bigl(\mathcal F_{[0,n]}(X)\bigr).
\]
Consequently,
\begin{equation}
\label{eq:shift-identity}
\begin{aligned}
\varphi(\theta(X))
&=(\omega\circ\mathcal E)\bigl(\mathcal F_{[0,n]}(X)\bigr),\\
\varphi(X)
&=\omega\bigl(\mathcal F_{[0,n]}(X)\bigr).
\end{aligned}
\end{equation}
These expressions agree for every local $X$ precisely when
$\omega\circ\mathcal E$ and $\omega$ agree on
$\mathcal R_{\mathcal F}$. Density of the local algebras
proves (b).

Part (c) follows from (b), since
$\mathcal F_\partial(\mathcal A_H)\subseteq\mathcal R_{\mathcal F}$.
Finally, (d) holds because the channel $\mathcal E_*$ has an
invariant density operator (Section~\ref{subsec:states}).
\end{proof}

The converse in (c) can fail without the range condition.

\begin{example}[Stationarity without an invariant boundary state]
\label{ex:stationarity-counterexample}
Let $d_H\geq2$, let $\gamma$ be a state on
$\mathcal A_H\otimes\mathcal A_O$, and let $\lambda$ be
a state on $\mathcal A_H$. Define
\[
\mathcal F(A\otimes x)=\gamma(A)\lambda(x)I_H.
\]
This map is completely positive and unital, and
\[
\mathcal F_\partial(a)=\gamma(a\otimes I_O)I_H,
\qquad
\mathcal E(x)=\lambda(x)I_H.
\]
For every $n\geq0$, the recursion gives
\[
\begin{aligned}
&\mathcal F_{[0,n]}
(a_0\otimes b_1\otimes a_1\otimes\cdots
 \otimes b_n\otimes a_n)\\
&\qquad=
\left(\prod_{k=0}^{n-1}\gamma(a_k\otimes b_{k+1})\right)
\gamma(a_n\otimes I_O)I_H,
\end{aligned}
\]
with the empty product equal to $1$.
Hence every boundary state generates the same stationary
product state, with state $\gamma$ on each block
$\mathcal A_k^H\otimes\mathcal A_{k+1}^O$.
However, $\omega\circ\mathcal E=\lambda$, so any
$\omega\neq\lambda$ is not invariant.
Here $\mathcal R_{\mathcal F}=\mathbb C I_H$.
\end{example}

Proposition~\ref{prop:stationarity} gives a finite test. Put
$R_0=\mathcal F_\partial(\mathcal A_H)$ and
$R_{n+1}=\operatorname{span}\{\mathcal F(A\otimes x):
A\in\mathcal A_H\otimes\mathcal A_O,\ x\in R_n\}$. Then
$R_n=\mathcal F_{[0,n]}(\mathcal A_{[0,n]})$; these subspaces increase
and stabilize in dimension at most $d_H^2$, and once $R_{n+1}=R_n$ they
remain constant and equal $\mathcal R_{\mathcal F}$.

\subsection{The hidden and observable marginals}
\label{subsec:hidden-marginal}
\label{subsec:observable-marginal}

Both marginals are obtained by inserting identities in the omitted
factors. Let
$\mathcal A^H_{[0,n]}=\bigotimes_{k=0}^{n}\mathcal A_k^H$ ($n\geq0$) and
$\mathcal A^O_{[1,n]}=\bigotimes_{k=1}^{n}\mathcal A_k^O$ ($n\geq1$),
with quasi-local algebras $\mathcal A^H_\infty$ and
$\mathcal A^O_\infty$ for the identifications $X\mapsto X\otimes I_H$
and $Y\mapsto Y\otimes I_O$, and define the unital injective
$*$-homomorphisms
\begin{equation}
\label{eq:hidden-embedding}
\begin{aligned}
\jmath_n:\mathcal A^H_{[0,n]}\to\mathcal A_{[0,n]},&\qquad
\jmath_n(a_0\otimes\cdots\otimes a_n)
=a_0\otimes I_O\otimes a_1\otimes\cdots\otimes I_O\otimes a_n,\\
\kappa_n:\mathcal A^O_{[1,n]}\to\mathcal A_{[0,n]},&\qquad
\kappa_n(b_1\otimes\cdots\otimes b_n)
=I_H\otimes b_1\otimes I_H\otimes\cdots\otimes b_n\otimes I_H.
\end{aligned}
\end{equation}
Since $\jmath_{n+1}(X\otimes I_H)=\iota_n(\jmath_n(X))$ and
$\kappa_{n+1}(Y\otimes I_O)=\iota_n(\kappa_n(Y))$, they extend to
unital injective $*$-homomorphisms
$\jmath_\infty:\mathcal A^H_\infty\to\mathcal A_\infty$ and
$\kappa_\infty:\mathcal A^O_\infty\to\mathcal A_\infty$. The states
$\varphi_H:=\varphi\circ\jmath_\infty$ and
$\varphi_O:=\varphi\circ\kappa_\infty$ are the \emph{hidden} and the
\emph{observable marginal} of $\varphi$. They are governed by the
\emph{hidden transition expectation} and the \emph{emission transition
expectation},
\begin{align}
\label{eq:hidden-te}
E_H^{\mathcal F}:\mathcal A_H\otimes\mathcal A_H\to\mathcal A_H,&\qquad
E_H^{\mathcal F}(a\otimes x)
=\mathcal F(a\otimes I_O\otimes x)
=\mathcal E_a(x),\\
\label{eq:emission-te}
\mathcal G:\mathcal A_O\otimes\mathcal A_H\to\mathcal A_H,&\qquad
\mathcal G(b\otimes x)
=\mathcal F(I_H\otimes b\otimes x).
\end{align}
Both are CP and unital, being compositions of $\mathcal F$ with unital
$*$-homomorphisms, and $\mathcal G(I_O\otimes x)=\mathcal E(x)$.

\begin{theorem}
\label{thm:hidden-marginal}
The state $\varphi_H$ is the quantum Markov chain generated
by $(\omega,E_H^{\mathcal F})$ in the sense of
Definition~\ref{def:qmc}. For every $n\geq0$ and
$a_0,\ldots,a_n\in\mathcal A_H$,
\begin{equation}
\label{eq:hidden-marginal-formula}
\varphi_H(a_0\otimes\cdots\otimes a_n)
=\omega\bigl(
\mathcal E_{a_0}\circ\cdots\circ\mathcal E_{a_n}(I_H)
\bigr).
\end{equation}
\end{theorem}

\begin{proof}
With the iterates $(E_H^{\mathcal F})_{[0,n]}$ of
\eqref{eq:qmc-iteration}, we have
$\mathcal F_{[0,n]}\circ\jmath_n=(E_H^{\mathcal F})_{[0,n]}$: for $n=0$
both sides are $\mathcal F_\partial(a)=E_H^{\mathcal F}(a\otimes I_H)$,
and the induction step is
\[
\begin{aligned}
\mathcal F_{[0,n]}(\jmath_n(a_0\otimes Y))
&=\mathcal F\bigl(a_0\otimes I_O\otimes
\mathcal F_{[0,n-1]}(\jmath_{n-1}(Y))\bigr)\\
&=E_H^{\mathcal F}\bigl(a_0\otimes(E_H^{\mathcal F})_{[0,n-1]}(Y)\bigr)
=(E_H^{\mathcal F})_{[0,n]}(a_0\otimes Y).
\end{aligned}
\]
Applying $\omega$ identifies $\varphi_H$ with the quantum Markov chain on
every local algebra, hence on $\mathcal A^H_\infty$, and
\eqref{eq:hidden-marginal-formula} follows from
$E_H^{\mathcal F}(a\otimes x)=\mathcal E_a(x)$.
\end{proof}

For $a\in\mathcal A_H$ and $k\geq0$, let $a^{(k)}$
denote the observable acting as $a$ at hidden site $k$
and as the identity elsewhere.

\begin{proposition}
\label{prop:hidden-correlations}
Let $r\geq1$, $0\leq k_1<\cdots<k_r$, and
$a_1,\ldots,a_r\in\mathcal A_H$. Then
\begin{equation}
\label{eq:hidden-correlation}
\begin{aligned}
&\varphi\bigl(a_1^{(k_1)}\cdots a_r^{(k_r)}\bigr)\\
&\quad=
\omega\Bigl(
\mathcal E^{k_1}\circ\mathcal E_{a_1}
\circ\mathcal E^{k_2-k_1-1}\circ\mathcal E_{a_2}
\circ\cdots\circ
\mathcal E^{k_r-k_{r-1}-1}\circ\mathcal E_{a_r}(I_H)
\Bigr),
\end{aligned}
\end{equation}
where, for $r=1$, the composition is
$\mathcal E^{k_1}\circ\mathcal E_{a_1}$.
In particular,
\begin{equation}
\label{eq:hidden-one-point}
\varphi(a^{(k)})
=\omega\bigl(\mathcal E^k(\mathcal F_\partial(a))\bigr),
\qquad k\geq0,
\end{equation}
and
\begin{equation}
\label{eq:hidden-two-point}
\varphi(a^{(j)}c^{(k)})
=\omega\Bigl(
\mathcal E^j\circ\mathcal E_a\circ\mathcal E^{k-j-1}
\bigl(\mathcal F_\partial(c)\bigr)
\Bigr),
\qquad 0\leq j<k.
\end{equation}
\end{proposition}

\begin{proof}
Apply \eqref{eq:hidden-marginal-formula} with $n=k_r$,
placing $a_i$ at site $k_i$ and the identity at every
other hidden site. Each identity contributes the map
$\mathcal E_{I_H}=\mathcal E$. There are $k_1$ such sites
before $k_1$ and $k_{i+1}-k_i-1$ between consecutive
specified sites. This gives \eqref{eq:hidden-correlation}.
The last two formulas follow from
$\mathcal E_a(I_H)=\mathcal F_\partial(a)$.
\end{proof}

\begin{theorem}
\label{thm:observable-marginal-fcs}
The state $\varphi_O$ is the finitely correlated state
generated by $(\omega,\mathcal G)$, with physical algebra
$\mathcal A_O$ and auxiliary algebra $\mathcal A_H$, in
the sense of Remark~\ref{rem:fcs-general}. For every
$n\geq1$ and $b_1,\ldots,b_n\in\mathcal A_O$,
\begin{equation}
\label{eq:observable-marginal-formula}
\varphi_O(b_1\otimes\cdots\otimes b_n)
=\omega\Bigl(
\mathcal G\bigl(b_1\otimes
\mathcal G(b_2\otimes\cdots
\mathcal G(b_n\otimes I_H)\cdots)\bigr)
\Bigr).
\end{equation}
Each finite observable marginal admits an open-boundary
matrix-product operator representation with bond dimension
at most $d_H^2$. If $\omega\circ\mathcal E=\omega$, then
$\varphi_O(I_O\otimes Y)=\varphi_O(Y)$ for every
$Y\in\mathcal A^O_{[1,n]}$ and $n\geq1$.
\end{theorem}

\begin{proof}
Insert $I_H$ at every hidden site in the recursion
\eqref{eq:mealy-iteration}: the terminal term is
$\mathcal F_\partial(I_H)=I_H$, and each preceding step is
$\mathcal F(I_H\otimes b\otimes x)=\mathcal G(b\otimes x)$. This gives
\eqref{eq:observable-marginal-formula}, the defining recursion of the
finitely correlated state generated by $(\omega,\mathcal G)$.

To prove the bond-dimension bound, choose a basis of
$\mathcal A_H$, whose complex dimension is $d_H^2$,
and matrix units $e_{uv}=|u\rangle\langle v|$ of
$\mathcal A_O$. Let $M_{uv}$ represent the linear map
\[
x\longmapsto\mathcal G(e_{uv}\otimes x)
\]
in the chosen basis of $\mathcal A_H$. Let $\ell$ be
the row vector representing $\omega$ and $r$ the column
vector representing $I_H$.

If $\rho_n$ is the density operator of
$\varphi_O|_{\mathcal A^O_{[1,n]}}$, then
\[
\begin{aligned}
\langle v_1\cdots v_n|\rho_n|u_1\cdots u_n\rangle
&=\varphi_O(e_{u_1v_1}\otimes\cdots\otimes e_{u_nv_n})\\
&=\ell M_{u_1v_1}\cdots M_{u_nv_n}r.
\end{aligned}
\]
Consequently,
\[
\rho_n
=\sum_{\boldsymbol u,\boldsymbol v}
\bigl(\ell M_{u_1v_1}\cdots M_{u_nv_n}r\bigr)
e_{v_1u_1}\otimes\cdots\otimes e_{v_nu_n}.
\]
Since each $M_{uv}$ has size $d_H^2\times d_H^2$,
this is the required representation.
If $\omega\circ\mathcal E=\omega$, then $\varphi\circ\theta=\varphi$ by
Proposition~\ref{prop:stationarity}(c), and since
$\theta(\kappa_n(Y))=\kappa_{n+1}(I_O\otimes Y)$,
\[
\varphi_O(I_O\otimes Y)
=\varphi(\theta(\kappa_n(Y)))
=\varphi(\kappa_n(Y))
=\varphi_O(Y),
\qquad Y\in\mathcal A^O_{[1,n]}.
\]
\end{proof}

\section{Moore Factorization of Mealy Transition Expectations}
\label{sec:moore-as-mealy}

Composing an emission map with a hidden transition map gives
a Mealy transition expectation. We identify the resulting
process with the Moore process and then ask which Mealy
transition expectations admit such a factorization on the
same hidden algebra.

\subsection{Moore factorization and the associated process}
\label{subsec:moore-recall}

Let
\begin{equation}
\label{eq:moore-data}
E_O:\mathcal A_H\otimes\mathcal A_O\longrightarrow\mathcal A_H,
\qquad
E_H:\mathcal A_H\otimes\mathcal A_H\longrightarrow\mathcal A_H
\end{equation}
be completely positive unital maps. Then
\begin{equation}
\label{eq:TM}
\mathcal F_{\mathrm M}
:=E_H\circ
\bigl(E_O\otimes\operatorname{id}_{\mathcal A_H}\bigr)
\end{equation}
is completely positive and unital, hence a Mealy transition
expectation. We call \eqref{eq:TM} a \emph{Moore factorization}
on $\mathcal A_H$.

For $a\in\mathcal A_H$ and $b\in\mathcal A_O$, put
\begin{equation}
\label{eq:moore-block}
\Gamma_{a,b}(x)
:=E_H\bigl(E_O(a\otimes b)\otimes x\bigr).
\end{equation}
In particular,
$\Gamma_{I_H,I_O}(I_H)=I_H$; the map
$\Gamma_{I_H,I_O}$ need not be the identity.

We index the observable emitted from hidden site $k$ by
$k+1$. Thus the Moore blocks are
$\mathcal A_k^H\otimes\mathcal A_{k+1}^O$.
Their local algebras
\[
\mathcal B_{[0,n]}
:=\bigotimes_{k=0}^{n}
\bigl(\mathcal A_k^H\otimes\mathcal A_{k+1}^O\bigr)
\]
and the Mealy local algebras satisfy
\[
\mathcal A_{[0,n]}
\subseteq \mathcal B_{[0,n]}
\subseteq \mathcal A_{[0,n+1]},
\]
where each inclusion appends the identity in the missing
factor. Their increasing unions therefore have the same
norm closure, namely $\mathcal A_\infty$.

Define
$M_{[0,n]}:\mathcal B_{[0,n]}\to\mathcal A_H$
recursively by
\begin{equation}
\label{eq:moore-iteration}
\begin{aligned}
M_{[0,0]}(A)
&=E_H\bigl(E_O(A)\otimes I_H\bigr),
\\
M_{[0,n]}
&=E_H\circ\bigl(E_O\otimes M_{[0,n-1]}\bigr),
\qquad n\geq1,
\end{aligned}
\end{equation}
with the factors of the tail relabelled in the second line.
Each $M_{[0,n]}$ is completely positive and unital,
by induction. Moreover,
$\Gamma_{I_H,I_O}(I_H)=I_H$ gives
\[
M_{[0,n+1]}(X\otimes I_H\otimes I_O)
=M_{[0,n]}(X).
\]
Thus the states $\omega\circ M_{[0,n]}$ are compatible
and extend uniquely to a state on $\mathcal A_\infty$.

\begin{definition}
\label{def:moore-process}
The state determined by
$\varphi^{\mathrm M}|_{\mathcal B_{[0,n]}}
=\omega\circ M_{[0,n]}$
is called the \emph{Moore process} generated by
$(\omega,E_O,E_H)$.
\end{definition}

On elementary tensors, its expectations are
\begin{equation}
\label{eq:moore-iteration-tensors}
\begin{aligned}
&\varphi^{\mathrm M}
\bigl(a_0\otimes b_1\otimes\cdots
      \otimes a_n\otimes b_{n+1}\bigr)
\\
&\qquad=
\omega\bigl(
\Gamma_{a_0,b_1}\circ\cdots\circ
\Gamma_{a_n,b_{n+1}}(I_H)
\bigr).
\end{aligned}
\end{equation}

\begin{theorem}
\label{thm:moore-mealy-identification}
The Mealy process generated by
$(\omega,\mathcal F_{\mathrm M})$ is
$\varphi^{\mathrm M}$. Its local expectations are
\begin{equation}
\label{eq:moore-mealy-finite-volume}
\begin{aligned}
&\varphi^{\mathrm M}
\bigl(a_0\otimes b_1\otimes a_1\otimes\cdots
      \otimes b_n\otimes a_n\bigr)
\\
&\qquad=
\omega\bigl(
\Gamma_{a_0,b_1}\circ\cdots\circ
\Gamma_{a_{n-1},b_n}\circ
\Gamma_{a_n,I_O}(I_H)
\bigr).
\end{aligned}
\end{equation}
For $n=0$, this expression is
$\omega(\Gamma_{a_0,I_O}(I_H))$.

Consequently, the Moore processes are exactly the Mealy
processes admitting a generating transition expectation
with a Moore factorization on the same hidden algebra.
\end{theorem}

\begin{proof}
The boundary map of $\mathcal F_{\mathrm M}$ is
\[
(\mathcal F_{\mathrm M})_\partial(a)
=\Gamma_{a,I_O}(I_H).
\]
Since
$\mathcal F_{\mathrm M}(a\otimes b\otimes x)
=\Gamma_{a,b}(x)$, the Mealy recursion
\eqref{eq:mealy-iteration} gives the right-hand side of
\eqref{eq:moore-mealy-finite-volume}.
The Moore formula \eqref{eq:moore-iteration-tensors}
gives the same expression upon setting $b_{n+1}=I_O$.
Thus the states agree on every $\mathcal A_{[0,n]}$
and, by continuity, on $\mathcal A_\infty$.

Every pair $(E_O,E_H)$ defines
$\mathcal F_{\mathrm M}$ by \eqref{eq:TM}.
Conversely, any generator with such a factorization
produces the Moore process associated with its factors.
\end{proof}

The factorization is a property of the chosen generator.
Its failure for one generator does not exclude a Moore
representation of the same state through another generator.
\subsection{A rank obstruction and its isometric characterization}
\label{subsec:moore-strict}

A Moore factorization imposes a rank bound on the transition
expectation. We establish this bound below and subsequently
prove that it characterizes Moore factorization for isometric
generators.

\begin{definition}
\label{def:lambda-b}
For a Mealy transition expectation $\mathcal F$, define the
linear map
\[
L_{\mathcal F}:
\mathcal A_H\otimes\mathcal A_O
\longrightarrow \mathcal L(\mathcal A_H),
\qquad
L_{\mathcal F}(A)(x)=\mathcal F(A\otimes x),
\]
where $\mathcal L(\mathcal A_H)$ denotes the vector space of
linear maps from $\mathcal A_H$ to itself. For elementary
tensors, write
\begin{equation}
\label{eq:lambda-ab}
\Lambda_{a,b}:=L_{\mathcal F}(a\otimes b),
\qquad
\Lambda_{a,b}(x)=\mathcal F(a\otimes b\otimes x).
\end{equation}
We associate with $\mathcal F$ the two ranks:
\begin{equation}
\label{eq:output-bond-rank}
\begin{aligned}
r_\star(\mathcal F)
&:=\operatorname{rank}L_{\mathcal F}
 =\dim_{\mathbb C}\operatorname{span}
   \{\Lambda_{a,b}:a\in\mathcal A_H,\ b\in\mathcal A_O\},
\\
r(\mathcal F)
&:=\operatorname{rank}
   \bigl(b\mapsto L_{\mathcal F}(I_H\otimes b)\bigr)
 =\dim_{\mathbb C}\operatorname{span}
   \{\Lambda_{I_H,b}:b\in\mathcal A_O\}.
\end{aligned}
\end{equation}
The expressions in terms of spans follow because elementary
tensors span $\mathcal A_H\otimes\mathcal A_O$.
\end{definition}

These ranks are independent of the choice of bases.
The dimensions of the domains and codomain give
\[
r(\mathcal F)\leq r_\star(\mathcal F)
\leq \min\{d_H^2d_O^2,d_H^4\},
\qquad
r(\mathcal F)\leq d_O^2.
\]

\begin{proposition}
\label{prop:rank-obstruction}
If $\mathcal F$ admits a Moore factorization on
$\mathcal A_H$, then
\begin{equation}
\label{eq:rank-bound}
r_\star(\mathcal F)\leq d_H^2.
\end{equation}
\end{proposition}

\begin{proof}
Suppose
$\mathcal F=E_H\circ
(E_O\otimes\operatorname{id}_{\mathcal A_H})$.
Define
\[
\Theta:\mathcal A_H\longrightarrow\mathcal L(\mathcal A_H),
\qquad
\Theta(y)(x)=E_H(y\otimes x).
\]
Then $L_{\mathcal F}=\Theta\circ E_O$, so
\[
r_\star(\mathcal F)
=\operatorname{rank}(\Theta\circ E_O)
\leq\dim_{\mathbb C}\mathcal A_H
=d_H^2.
\]
\end{proof}

\subsubsection{Block decomposition and the rank formula}
\label{subsubsec:block-form}

For an isometric transition expectation, the rank
$r_\star$ can be computed from the blocks of its implementing
isometry. The same decomposition will be used in
Section~\ref{sec:observable-reduction}.

Fix orthonormal bases
$\{|i\rangle\}_{i=1}^{d_H}$ of $\mathcal H_H$ and
$\{|m\rangle\}_{m=1}^{d_O}$ of $\mathcal H_O$,
and put $b_m=|m\rangle\langle m|$.

\begin{lemma}
\label{lem:block-form}
Every linear map
$W:\mathcal H_H\to
\mathcal H_H\otimes\mathcal H_O\otimes\mathcal H_H$
has a unique decomposition
\begin{equation}
\label{eq:block-form}
W\psi
=\sum_{i=1}^{d_H}\sum_{m=1}^{d_O}
|i\rangle\otimes|m\rangle\otimes W_{i,m}\psi,
\qquad
W_{i,m}\in\mathcal B(\mathcal H_H).
\end{equation}
Moreover, $W$ is an isometry if and only if
\begin{equation}
\label{eq:block-isometry}
\sum_{i,m}W_{i,m}^*W_{i,m}=I_H.
\end{equation}

For $\mathcal F=\operatorname{Ad}_W$, with
$\operatorname{Ad}_W(Z)=W^*ZW$, one has
\begin{equation}
\label{eq:block-general}
\mathcal F(a\otimes y\otimes x)
=\sum_{i,j=1}^{d_H}\sum_{m,m'=1}^{d_O}
\langle j|a|i\rangle
\langle m'|y|m\rangle
W_{j,m'}^*xW_{i,m}.
\end{equation}
In particular,
\begin{equation}
\label{eq:block-reduction}
\Phi_m(x):=\mathcal F(I_H\otimes b_m\otimes x)
=\sum_i W_{i,m}^*xW_{i,m},
\end{equation}
and
\begin{equation}
\label{eq:block-slot}
\mathcal E_a(x)
=\sum_m\sum_{i,j}
\langle j|a|i\rangle W_{j,m}^*xW_{i,m}.
\end{equation}
\end{lemma}

\begin{proof}
The blocks are necessarily
\[
W_{i,m}
=(\langle i|\otimes\langle m|\otimes I_H)W,
\]
which also proves their existence and uniqueness.
Orthonormality gives
\[
W^*W=\sum_{i,m}W_{i,m}^*W_{i,m},
\]
and hence \eqref{eq:block-isometry}.
Expanding $W^*(a\otimes y\otimes x)W$ gives
\eqref{eq:block-general}.
Taking $(a,y)=(I_H,b_m)$ gives
\eqref{eq:block-reduction}, while taking $y=I_O$
gives \eqref{eq:block-slot}.
\end{proof}

The rank formula uses the two-sided multiplication map.

\begin{lemma}
\label{lem:mult-iso}
The linear map
$\mu:\mathcal B(\mathcal H)\otimes\mathcal B(\mathcal H)
\to\mathcal L(\mathcal B(\mathcal H))$,
$\mu(A\otimes B)(x)=AxB$, is an isomorphism. Consequently, for every
linear subspace $\mathcal V\subseteq\mathcal B(\mathcal H)$,
\[
\dim_{\mathbb C}\operatorname{span}
\{x\mapsto A^*xB:A,B\in\mathcal V\}
=(\dim_{\mathbb C}\mathcal V)^2.
\]
\end{lemma}

\begin{proof}
For an orthonormal basis $(|j\rangle)_j$ and
$\operatorname{vec}(x)=\sum_j x|j\rangle\otimes|j\rangle$, one has
$\operatorname{vec}(AxB)=(A\otimes B^{\mathsf T})\operatorname{vec}(x)$.
Thus $\mu$ is the invertible map
$\operatorname{id}\otimes(\cdot)^{\mathsf T}$ followed by the canonical
identification of the tensor-product matrix algebra with
$\mathcal B(\mathcal H\otimes\mathcal H)$. The span in the second
assertion is $\mu(\mathcal V^*\otimes\mathcal V)$, where
$\mathcal V^*=\{A^*:A\in\mathcal V\}$ has the same dimension as
$\mathcal V$.
\end{proof}

\begin{proposition}
\label{prop:rstar-block}
Let $\mathcal F=\operatorname{Ad}_W$ be an isometric
transition expectation, and set
\begin{equation}
\label{eq:block-span}
\mathcal V_W
:=\operatorname{span}
\{W_{i,m}:1\leq i\leq d_H,\ 1\leq m \leq d_O\}.
\end{equation}
Then
\begin{equation}
\label{eq:rstar-block}
r_\star(\mathcal F)
=\bigl(\dim_{\mathbb C}\mathcal V_W\bigr)^2.
\end{equation}
\end{proposition}

\begin{proof}
By \eqref{eq:block-general}, every $\Lambda_{a,b}$
is a linear combination of the maps
$x\mapsto W_{j,m'}^*xW_{i,m}$.
Conversely, taking
$a=|j\rangle\langle i|$ and
$b=|m'\rangle\langle m|$ gives
\[
\Lambda_{a,b}(x)=W_{j,m'}^*xW_{i,m}.
\]
Since the blocks span $\mathcal V_W$, it follows that
\[
\operatorname{ran}L_{\mathcal F}
=\operatorname{span}
\{x\mapsto A^*xB:A,B\in\mathcal V_W\}.
\]
By Lemma~\ref{lem:mult-iso}, this space
has dimension $(\dim_{\mathbb C}\mathcal V_W)^2$,
which proves \eqref{eq:rstar-block}.
\end{proof}
\subsubsection{Characterization of isometric Moore factorizations}

\begin{theorem}
\label{thm:isometric-moore-characterization}
Let
$W:\mathcal H_H\to
\mathcal H_H\otimes\mathcal H_O\otimes\mathcal H_H$
be an isometry. Set
$\mathcal F=\operatorname{Ad}_W$ and
$r_W=\dim_{\mathbb C}\mathcal V_W$.
The following assertions are equivalent:
\begin{enumerate}
\item[(i)]
$\mathcal F$ admits a Moore factorization on
$\mathcal A_H$ by completely positive unital maps.

\item[(ii)]
$r_\star(\mathcal F)\leq d_H^2$.

\item[(iii)]
$r_W\leq d_H$.

\item[(iv)]
There exist isometries
\[
V_O:\mathcal H_H\longrightarrow
\mathcal H_H\otimes\mathcal H_O,
\qquad
V_H:\mathcal H_H\longrightarrow
\mathcal H_H\otimes\mathcal H_H
\]
such that
\[
W=(V_O\otimes I_H)V_H.
\]
\end{enumerate}
Whenever these conditions hold, a Moore factorization
can be chosen with
$E_O=\operatorname{Ad}_{V_O}$ and
$E_H=\operatorname{Ad}_{V_H}$.
\end{theorem}

\begin{proof}
Proposition~\ref{prop:rank-obstruction} gives
{\rm(i)}$\Rightarrow${\rm(ii)}, and
\eqref{eq:rstar-block} gives
{\rm(ii)}$\Leftrightarrow${\rm(iii)}.

Suppose {\rm(iii)} holds.
Regard $W$ as an element of
\[
(\mathcal H_H\otimes\mathcal H_O)
\otimes\mathcal B(\mathcal H_H),
\]
where $\mathcal B(\mathcal H_H)$ carries the
Hilbert--Schmidt inner product.
Its Schmidt rank is the dimension of the span of
its blocks, namely $r_W$. Absorbing the nonzero Schmidt
coefficients into the matrix factors gives
\begin{equation}
\label{eq:W-physical-schmidt}
W\psi
=\sum_{\alpha=1}^{r_W}
v_\alpha\otimes A_\alpha\psi,
\end{equation}
where the $v_\alpha$ are orthonormal and the
$A_\alpha$ are linearly independent.
Since $W$ is an isometry,
\begin{equation}
\label{eq:schmidt-normalization}
\sum_{\alpha=1}^{r_W}A_\alpha^*A_\alpha=I_H.
\end{equation}

Since
$r_W\leq d_H\leq
\dim(\mathcal H_H\otimes\mathcal H_O)=d_Hd_O$,
extend $(v_\alpha)_{\alpha=1}^{r_W}$ to an
orthonormal family of length $d_H$ in
$\mathcal H_H\otimes\mathcal H_O$.
For an orthonormal basis
$(|\alpha\rangle)_{\alpha=1}^{d_H}$ of $\mathcal H_H$,
define
\[
V_O|\alpha\rangle=v_\alpha,
\qquad
V_H\psi
=\sum_{\alpha=1}^{r_W}
|\alpha\rangle\otimes A_\alpha\psi.
\]
The map $V_O$ is an isometry by construction,
and $V_H$ is an isometry by
\eqref{eq:schmidt-normalization}.
Equation~\eqref{eq:W-physical-schmidt} then gives
$W=(V_O\otimes I_H)V_H$, proving {\rm(iv)}.

Finally, {\rm(iv)} implies
\[
\operatorname{Ad}_W
=\operatorname{Ad}_{V_H}\circ
\bigl(\operatorname{Ad}_{V_O}
      \otimes\operatorname{id}_{\mathcal A_H}\bigr).
\]
Both factors are completely positive and unital,
which proves {\rm(i)}.
\end{proof}

The factorization can be computed by a singular value
decomposition. Expand
$W_{i,m}=\sum_\beta C_{(i,m),\beta}B_\beta$
in a Hilbert--Schmidt orthonormal matrix basis
$(B_\beta)_\beta$.
The coefficient matrix $C$ has rank $r_W$,
and its left singular vectors corresponding to
nonzero singular values give the vectors $v_\alpha$
in \eqref{eq:W-physical-schmidt}.

\begin{corollary}
\label{cor:minimal-intermediate-dimension}
Let $W$ be as in
Theorem~\ref{thm:isometric-moore-characterization}.
Among factorizations
\begin{equation}
\label{eq:intermediate-factorization}
\operatorname{Ad}_W
=P\circ
\bigl(Q\otimes\operatorname{id}_{\mathcal A_H}\bigr),
\end{equation}
where
\[
Q:\mathcal A_H\otimes\mathcal A_O
\longrightarrow\mathcal B(\mathcal K),
\qquad
P:\mathcal B(\mathcal K)\otimes\mathcal A_H
\longrightarrow\mathcal A_H
\]
are completely positive unital maps, the smallest
possible dimension of $\mathcal K$ is $r_W$.
The minimum is attained by maps implemented by isometries.
\end{corollary}

\begin{proof}
Set $k=\dim\mathcal K$.
As in Proposition~\ref{prop:rank-obstruction},
the factorization \eqref{eq:intermediate-factorization}
makes $L_{\mathcal F}$ factor through
$\mathcal B(\mathcal K)$. Hence
\[
r_W^2=r_\star(\mathcal F)\leq k^2,
\]
so $r_W\leq k$.

For $\mathcal K=\mathbb C^{r_W}$, use
\eqref{eq:W-physical-schmidt} to define
\[
V_O:\mathcal K\longrightarrow
\mathcal H_H\otimes\mathcal H_O,
\qquad
V_O|\alpha\rangle=v_\alpha,
\]
and
\[
V_H:\mathcal H_H\longrightarrow
\mathcal K\otimes\mathcal H_H,
\qquad
V_H\psi
=\sum_{\alpha=1}^{r_W}
|\alpha\rangle\otimes A_\alpha\psi.
\]
These maps are isometries and satisfy
$W=(V_O\otimes I_H)V_H$.
Thus $Q=\operatorname{Ad}_{V_O}$ and
$P=\operatorname{Ad}_{V_H}$ attain the bound.
\end{proof}

Here only the intermediate space $\mathcal K$ varies;
the domain and codomain of $W$ remain fixed.
The corollary therefore does not assert minimality
of the memory dimension of the generated process.

For isometric generators, the block span is determined by the hidden
channel alone.

\begin{corollary}
\label{cor:factorization-channel}
Let $W$ be as in Theorem~\ref{thm:isometric-moore-characterization},
and let $\mathcal E$ be the hidden channel of $\operatorname{Ad}_W$.
Then $\mathcal V_W$ is the linear span of the operators of any Kraus
representation of $\mathcal E$, and $r_W$ is the Kraus rank of
$\mathcal E$, that is, the smallest number of operators in a Kraus
representation of $\mathcal E$. Consequently, $\operatorname{Ad}_W$
admits a Moore factorization on $\mathcal A_H$ if and only if
$\mathcal E$ has a Kraus representation with at most $d_H$ operators,
and the minimal intermediate dimension in
Corollary~\ref{cor:minimal-intermediate-dimension} is the Kraus rank of
$\mathcal E$. In particular, isometric generators with the same hidden
channel, for instance with the same observable reduction for some
projective readout (Section~\ref{sec:observable-reduction}), are either
all Moore-factorizable or none is.
\end{corollary}

\begin{proof}
By \eqref{eq:block-slot} with $a=I_H$,
$\mathcal E(x)=\sum_{i,m}W_{i,m}^*xW_{i,m}$, so the blocks form a
Kraus family of $\mathcal E$. Let
$|\Omega\rangle=\sum_k|k\rangle\otimes|k\rangle$. For any Kraus family
$(V_\alpha)_\alpha$ of $\mathcal E$, the Choi matrix of
$\mathcal E_*$ is
$J_{\mathcal E}=\sum_\alpha(I_H\otimes V_\alpha)|\Omega\rangle\langle\Omega|
(I_H\otimes V_\alpha)^*$ \cite{Choi1975}, whose range is
$\{(I_H\otimes A)|\Omega\rangle:A\in\operatorname{span}_\alpha\{V_\alpha\}\}$.
Since $A\mapsto(I_H\otimes A)|\Omega\rangle$ is injective, every Kraus
family of $\mathcal E$ spans the same space as the blocks, namely
$\mathcal V_W$, and $\operatorname{rank}J_{\mathcal E}=r_W$. A Kraus
family with $k$ operators gives $\operatorname{rank}J_{\mathcal E}\leq k$,
while the eigenvectors of $J_{\mathcal E}$ for its nonzero eigenvalues,
scaled by the square roots of these eigenvalues, give a Kraus family
with $\operatorname{rank}J_{\mathcal E}$ operators \cite{Choi1975};
hence $r_W$ is the Kraus rank of $\mathcal E$. The
remaining assertions follow from
Theorem~\ref{thm:isometric-moore-characterization} and
Corollary~\ref{cor:minimal-intermediate-dimension}, since
$\sum_m\mathcal F(I_H\otimes b_m\otimes x)=\mathcal E(x)$ for every
projective readout $\{b_m\}$.
\end{proof}

\subsubsection{Explicit obstructions to Moore factorization}
\label{subsubsec:obstructions}

\begin{example}[$d_H=2$, $d_O=3$]
\label{ex:non-moore}
Let $\mathcal H_H=\mathbb C^{2}$, $\mathcal H_O=\mathbb C^{3}$, put
$S_1=I$, $S_2=\sigma_x$, $S_3=\sigma_y$, and define
\begin{equation}
\label{eq:non-moore-isometry}
W\psi=\frac{1}{\sqrt3}\sum_{m=1}^{3}
|1\rangle\otimes|m\rangle\otimes S_m\psi .
\end{equation}
The blocks are $W_{1,m}=S_m/\sqrt3$ and $W_{2,m}=0$, so
$\sum_{i,m}W_{i,m}^{*}W_{i,m}=\frac13\sum_mS_m^{*}S_m=I$ and $W$ is an
isometry. Since $\{I,\sigma_x,\sigma_y\}$ is linearly independent in
$M_2(\mathbb C)$, we get $\dim\mathcal V_W=3>2=d_H$, hence $r_\star(\mathcal F)=9>4$ by Proposition~\ref{prop:rstar-block} and, by Proposition~\ref{prop:rank-obstruction}, $\mathcal F=\operatorname{Ad}_W$ admits no Moore factorization on $\mathcal A_H=M_2(\mathbb C)$. Here the site
leg is the fixed vector $|1\rangle$, so $r(\mathcal F)=r_\star(\mathcal F)=9$ and the
weaker invariant already suffices.
\end{example}

\begin{example}[Equal hidden and output dimensions]
\label{ex:equal-dimension-non-moore}
Let $d_H=d_O=2$, and in the block notation of
Lemma~\ref{lem:block-form} take
\[
W_{1,0}=\tfrac12 I,\qquad W_{1,1}=\tfrac12\sigma_x,\qquad
W_{2,0}=\tfrac12\sigma_y,\qquad W_{2,1}=\tfrac12\sigma_z.
\]
Here the output basis is indexed by $0,1$. Since each Pauli matrix
is unitary,
$\sum_{i,m}W_{i,m}^*W_{i,m}=I$; hence these blocks define an isometry
$W$. Moreover $\mathcal V_W=M_2(\mathbb C)$, so
\[
r_\star(\mathcal F)=16>4=d_H^2,\qquad \mathcal F=\operatorname{Ad}_W.
\]
Proposition~\ref{prop:rank-obstruction} excludes a Moore factorization
on the same hidden algebra.

For completeness, put $e_{uv}=|u\rangle\langle v|$. The four slices
$\Lambda_{I_H,e_{uv}}(x)=\mathcal F(I_H\otimes e_{uv}\otimes x)$ are
\[
\begin{aligned}
4\Lambda_{I_H,e_{00}}(x)&=x+\sigma_yx\sigma_y,&
4\Lambda_{I_H,e_{11}}(x)&=\sigma_xx\sigma_x+\sigma_zx\sigma_z,\\
4\Lambda_{I_H,e_{01}}(x)&=x\sigma_x+\sigma_yx\sigma_z,&
4\Lambda_{I_H,e_{10}}(x)&=\sigma_xx+\sigma_zx\sigma_y.
\end{aligned}
\]
Under the two-sided multiplication isomorphism of
Lemma~\ref{lem:mult-iso}, these maps have disjoint supports in the
Pauli tensor basis and are nonzero, hence linearly independent.
Therefore $r(\mathcal F)=4$: the weaker invariant does not detect
this obstruction. This example has equal hidden and output dimensions
and a nonconstant hidden-site component.
\end{example}

\subsubsection{Genericity of the obstruction}
\label{subsubsec:genericity}

\begin{theorem}
\label{thm:moore-nongeneric}
Let $\mathcal W$ be the set of isometries
$W:\mathcal H_H\to\mathcal H_H\otimes\mathcal H_O\otimes\mathcal H_H$,
and let $\mathcal W_{\mathrm M}\subseteq\mathcal W$ consist of those $W$
for which $\operatorname{Ad}_W$ admits a Moore factorization on
$\mathcal A_H$. If $d_H\geq2$ and $d_O\geq2$, then
$\mathcal W_{\mathrm M}$ is a closed real-algebraic subset of
$\mathcal W$ with empty interior, and it has measure zero for the
unitarily invariant probability measure on $\mathcal W$. If $d_H=1$ or
$d_O=1$, then $\mathcal W_{\mathrm M}=\mathcal W$.
\end{theorem}

\begin{proof}
By Theorem~\ref{thm:isometric-moore-characterization},
$W\in\mathcal W_{\mathrm M}$ if and only if $r_W\leq d_H$. If $d_H=1$,
then $\mathcal V_W\subseteq\mathcal B(\mathbb C)$; if $d_O=1$, there are
only $d_H$ blocks. In both cases $r_W\leq d_H$.

Let $d_H,d_O\geq2$. In a Hilbert--Schmidt orthonormal basis
$(B_\beta)_\beta$ of $\mathcal A_H$, the coefficient matrix
$C(W)=\bigl(\operatorname{Tr}(B_\beta^*W_{i,m})\bigr)_{(i,m),\beta}$
has rank $r_W$, and its entries are linear in the matrix entries of
$W$. Thus $\mathcal W_{\mathrm M}$ is the zero set in $\mathcal W$ of
the real polynomial $f(W)=\sum_\mu|\mu(W)|^2$, where $\mu$ runs over
the minors of order $d_H+1$ of $C(W)$. Since $\mathcal W$ is defined
by the polynomial equation $W^*W=I_H$, the set $\mathcal W_{\mathrm M}$
is closed and real-algebraic.

The polynomial $f$ does not vanish identically on $\mathcal W$. Since
$d_Hd_O\geq d_H+1$ and $d_H^2\geq d_H+1$, choose distinct index pairs
$(i_1,m_1),\ldots,(i_{d_H+1},m_{d_H+1})$ and distinct generalized Pauli
operators $U_1,\ldots,U_{d_H+1}$ of the form $X_H^aZ_H^b$
($0\leq a,b<d_H$), where $X_H|s\rangle=|s+1\rangle$ and
$Z_H|s\rangle=e^{2\pi\mathrm is/d_H}|s\rangle$ for an orthonormal basis
$(|s\rangle)_{s=0}^{d_H-1}$ of $\mathcal H_H$, indices being taken
modulo $d_H$. These unitaries are pairwise orthogonal for the
Hilbert--Schmidt inner product. The blocks
$W_{i_k,m_k}=(d_H+1)^{-1/2}U_k$, all other blocks being zero, satisfy
\eqref{eq:block-isometry} and span a space of dimension $d_H+1$.
Example~\ref{ex:equal-dimension-non-moore} is a construction of the
same kind with all four Pauli matrices.

Let $N=d_H^2d_O$ and fix $J\in\mathcal W$. Every isometry extends to a
unitary, so $U\mapsto UJ$ maps the unitary group $\mathcal U(N)$ onto
$\mathcal W$, and the unitarily invariant probability measure on
$\mathcal W$ is the image of the Haar measure. The function
$U\mapsto f(UJ)$ is real-analytic on the connected real-analytic
manifold $\mathcal U(N)$ and does not vanish identically, so its zero
set, the preimage of $\mathcal W_{\mathrm M}$, has Haar measure zero
\cite{Mityagin2020}. Hence $\mathcal W_{\mathrm M}$ has measure zero.
A nonempty open subset of $\mathcal W$ has a nonempty open preimage,
of positive Haar measure, so $\mathcal W_{\mathrm M}$ has empty
interior.
\end{proof}


\section{Operational HQMMs as Observable Reductions}
\label{sec:observable-reduction}

This section establishes the relation between Mealy hidden quantum
Markov processes and the operational hidden quantum Markov models
(HQMMs) of Monr\`as, Beige and Wiesner
\cite{MonrasBeigeWiesner2011}.

Throughout this section, let $\mathcal F$ be a Mealy transition
expectation, $\omega$ a boundary state, and $\varphi$ the associated
quasi-local state. Let $\mathcal G$ be the emission transition
expectation defined in \eqref{eq:emission-te}.
Fix a family $\mathsf b=\{b_m\}_{m\in O}$ in $\mathcal A_O$ with
\begin{equation}
\label{eq:readout}
b_m=b_m^{*}=b_m^{2}\neq0,
\qquad
b_mb_{m'}=0\ \ (m\neq m'),
\qquad
\sum_{m\in O}b_m=I_O .
\end{equation}
We call $\mathsf b$ the output \emph{readout} and denote by
\[
\mathcal D_{\mathsf b}
=\operatorname{span}\{b_m:m\in O\}\simeq C(O)
\]
the commutative unital subalgebra it generates.
The observable reduction is defined relative to $\mathsf b$.
From Section~\ref{subsec:lifting} onwards, $\mathsf b$ is the
rank-one readout attached to an orthonormal basis of
$\mathcal H_O$.

\subsection{The commutative reduction}
\label{subsec:commutative-reduction}

In the commutative setting, the reduction recovers the
output-labelled transition matrices of a classical Mealy model.

Let $\mathcal F_{\mathrm{cl}}$ be as in
\eqref{eq:classical-mealy-te}, and let $\chi_m\in C(O)$
be the indicator of $\{m\}$. Define
\begin{equation}
\label{eq:commutative-reduction}
\mathcal P_m:C(S)\to C(S),
\qquad
\mathcal P_m(h)
=\mathcal F_{\mathrm{cl}}
\bigl(\mathbf 1\otimes\chi_m\otimes h\bigr).
\end{equation}

\begin{proposition}
\label{prop:commutative-reduction}
For every $m\in O$, $h\in C(S)$, and $i\in S$,
\[
(\mathcal P_mh)(i)
=\sum_{j\in S}p(j,m\mid i)\,h(j).
\]
Thus $\mathcal P_m$ is represented by the output-labelled
matrix $P_m$ of \eqref{eq:output-labelled-matrix}.
Each $\mathcal P_m$ is positive and subunital, and
$\sum_{m\in O}\mathcal P_m$ is unital.
For every initial distribution $\pi$ and every word
$(m_1,\ldots,m_n)\in O^n$,
\begin{equation}
\label{eq:commutative-word}
\mathbb P_\pi(m_1,\ldots,m_n)
=\pi\bigl(
\mathcal P_{m_1}\circ\cdots\circ
\mathcal P_{m_n}(\mathbf 1)
\bigr)
=\pi P_{m_1}\cdots P_{m_n}\mathbf 1,
\end{equation}
where $\pi$ is identified with a state on $C(S)$
and with its probability row vector.
\end{proposition}

\begin{proof}
The formula for $\mathcal P_mh$ follows by substituting
$\chi_m(m')=\delta_{m,m'}$ into
\eqref{eq:classical-mealy-te}.
Since $p(j,m\mid i)\geq0$ and
$\sum_{m\in O}\sum_{j\in S}p(j,m\mid i)=1$,
each $\mathcal P_m$ is positive and
\[
0\leq\mathcal P_m(\mathbf 1)\leq\mathbf 1,
\qquad
\sum_{m\in O}\mathcal P_m(\mathbf 1)=\mathbf 1.
\]
The word-probability identity follows from
\eqref{eq:classical-word-probability}.
\end{proof}

\begin{proposition}
\label{prop:commutative-bijection}
The assignment
$\mathcal F_{\mathrm{cl}}\mapsto\{\mathcal P_m\}_{m\in O}$
of \eqref{eq:commutative-reduction} is a bijection from the
classical Mealy transition expectations of the form
\eqref{eq:classical-mealy-te} onto the families of positive maps
$\mathcal P_m:C(S)\to C(S)$ satisfying
$\sum_{m\in O}\mathcal P_m(\mathbf 1)=\mathbf 1$.
Its inverse is
\begin{equation}
\label{eq:commutative-inverse}
\mathcal F_{\mathrm{cl}}(f\otimes g\otimes h)
=f\sum_{m\in O}g(m)\,\mathcal P_m(h),
\qquad
f,h\in C(S),\quad g\in C(O),
\end{equation}
multiplication by $f$ being pointwise.
\end{proposition}

\begin{proof}
Proposition~\ref{prop:commutative-reduction} shows that the
assignment takes values in the stated set. It is injective:
taking $h=e_j$ in that proposition gives
$p(j,m\mid i)=(\mathcal P_me_j)(i)$, so $\{\mathcal P_m\}_{m\in O}$
determines the kernel, and the kernel determines
$\mathcal F_{\mathrm{cl}}$ through \eqref{eq:classical-mealy-te}.
For surjectivity, let $\{\mathcal P_m\}_{m\in O}$ be such a family
and put
\[
p(j,m\mid i)=(\mathcal P_me_j)(i),
\qquad
e_j=\mathbf 1_{\{j\}}\in C(S).
\]
Positivity of $\mathcal P_m$ gives $p(j,m\mid i)\geq0$, and
\[
\sum_{m\in O}\sum_{j\in S}p(j,m\mid i)
=\sum_{m\in O}\bigl(\mathcal P_m\mathbf 1\bigr)(i)=1 ,
\]
so $p$ is a Mealy kernel. The map $\mathcal F_{\mathrm{cl}}$ it
generates through \eqref{eq:classical-mealy-te} is positive and
unital, hence completely positive since the algebras are
commutative, and it satisfies \eqref{eq:commutative-inverse} on
elementary tensors, which span the domain. Its reduction is
$\{\mathcal P_m\}$ by Proposition~\ref{prop:commutative-reduction}.
\end{proof}

Injectivity uses the form \eqref{eq:classical-mealy-te}, in which the
hidden-site argument is evaluated at the current hidden state, that is,
copied. For a general positive unital map
$\mathcal F(\eta)(i)=\sum_{h,m,j}q(h,m,j\mid i)\,\eta(h,m,j)$ on the same
commutative algebras, the reduction retains only $\sum_hq(h,m,j\mid i)$,
so injectivity fails already classically. On a full matrix algebra, the copying identity forces the hidden channel to have scalar range
(Proposition~\ref{prop:copying-scalar}), and the reduction is surjective
but not injective, as shown below.

\subsection{Observable reduction of a Mealy transition expectation}
\label{subsec:observable-reduction}

\begin{definition}
\label{def:observable-reduction}
The \emph{observable reduction} of $\mathcal F$ relative to
$\mathsf b$ is the family of maps
\begin{equation}
\label{eq:observable-reduction}
\Phi_m:\mathcal A_H\to\mathcal A_H,
\qquad
\Phi_m(x)=\mathcal G(b_m\otimes x),
\qquad m\in O ;
\end{equation}
equivalently $\Phi_m(x)=\mathcal F(I_H\otimes b_m\otimes x)$, by
\eqref{eq:emission-te}.
Their Schr\"odinger duals
$\mathcal K_m=(\Phi_m)_*$ are determined by
\begin{equation}
\label{eq:observable-reduction-duality}
\operatorname{Tr}\bigl(\mathcal K_m(\rho)x\bigr)
=\operatorname{Tr}\bigl(\rho\,\Phi_m(x)\bigr),
\qquad
\rho\in\mathcal T(\mathcal H_H),\quad x\in\mathcal A_H.
\end{equation}
We denote the reduction in the Schr\"odinger picture by
\[
R_{\mathsf b}(\mathcal F)
=\{\mathcal K_m\}_{m\in O}.
\]
\end{definition}
The reduction is defined on generators, not on process states: for the
generators of Example~\ref{ex:stationarity-counterexample},
$\Phi_m(x)=\gamma(I_H\otimes b_m)\lambda(x)I_H$ depends on $\lambda$,
whereas the generated state does not.

\begin{proposition}
\label{prop:reduced-hqmm}
Each $\Phi_m$ is completely positive and subunital, and
\begin{equation}
\label{eq:reduction-completeness}
\sum_{m\in O}\Phi_m(I_H)=I_H.
\end{equation}
Consequently, each $\mathcal K_m$ is completely positive and
trace non-increasing, and $\sum_{m\in O}\mathcal K_m$ is trace
preserving. Thus $\{\mathcal K_m\}_{m\in O}$ is a finite-outcome
quantum instrument in the sense of Definition~\ref{def:hqmm}.
\end{proposition}

\begin{proof}
Since $b_m$ is a projection, the map $x\mapsto b_m\otimes x$
is a $*$-homomorphism, hence completely positive.
Its composition with $\mathcal G$ is $\Phi_m$, so $\Phi_m$
is completely positive. Moreover,
\[
\sum_{m\in O}\Phi_m(I_H)
=\mathcal G\Bigl(\sum_{m\in O}b_m\otimes I_H\Bigr)
=\mathcal G(I_O\otimes I_H)
=I_H ,
\]
by \eqref{eq:readout} and unitality of $\mathcal G$.
Since each summand is positive, $\Phi_m(I_H)\leq I_H$.

Complete positivity passes to the Schr\"odinger duals.
For every $\rho\geq0$, duality gives
\[
\operatorname{Tr}\bigl(\mathcal K_m(\rho)\bigr)
=\operatorname{Tr}\bigl(\rho\,\Phi_m(I_H)\bigr)
\leq\operatorname{Tr}\rho.
\]
Summing over $m$ and using
\eqref{eq:reduction-completeness} yields
\[
\operatorname{Tr}\Bigl(\sum_{m\in O}\mathcal K_m(\rho)\Bigr)
=\operatorname{Tr}\rho.
\]
\end{proof}

Let $\rho_\omega$ be the density operator of $\omega$.
For $w=(m_1,\ldots,m_n)\in O^n$, define the cylinder projection
\begin{equation}
\label{eq:cylinder}
B_w
=I_H\otimes b_{m_1}\otimes I_H\otimes\cdots
\otimes b_{m_n}\otimes I_H
\in\mathcal A_{[0,n]}.
\end{equation}

\begin{theorem}
\label{thm:observable-reduction}
For every $n\geq1$ and $w=(m_1,\ldots,m_n)\in O^n$,
\begin{equation}
\label{eq:observable-reduction-word}
\varphi(B_w)
=\omega\bigl(
\Phi_{m_1}\circ\cdots\circ\Phi_{m_n}(I_H)
\bigr)
=\operatorname{Tr}\bigl[
\mathcal K_{m_n}\circ\cdots\circ
\mathcal K_{m_1}(\rho_\omega)
\bigr].
\end{equation}
Hence the restriction of $\varphi$ to the commutative output
algebra determined by $\mathsf b$ coincides with the output law
of the operational HQMM
$\bigl(\mathcal H_H,\rho_\omega,
\{\mathcal K_m\}_{m\in O}\bigr)$.
\end{theorem}

\begin{proof}
We prove by induction on $n$ that
\begin{equation}
\label{eq:cylinder-iterated}
\mathcal F_{[0,n]}(B_w)
=\Phi_{m_1}\circ\cdots\circ\Phi_{m_n}(I_H).
\end{equation}
For $n=1$, the recursion \eqref{eq:mealy-iteration} and
\eqref{eq:T0-unital} give
\[
\mathcal F_{[0,1]}(B_w)
=\mathcal F\bigl(
I_H\otimes b_{m_1}\otimes\mathcal F_\partial(I_H)
\bigr)
=\Phi_{m_1}(I_H).
\]
Suppose the identity holds for words of length $n$.
Let $w=(m_1,\ldots,m_{n+1})$ and
$w'=(m_2,\ldots,m_{n+1})$.
Relabelling the factors that follow
$\mathcal A_0^H\otimes\mathcal A_1^O$ as in
\eqref{eq:mealy-iteration} gives
$B_w=I_H\otimes b_{m_1}\otimes B_{w'}$, whence
\[
\begin{aligned}
\mathcal F_{[0,n+1]}(B_w)
&=\mathcal F\bigl(
I_H\otimes b_{m_1}\otimes
\mathcal F_{[0,n]}(B_{w'})
\bigr)\\
&=\Phi_{m_1}\bigl(\mathcal F_{[0,n]}(B_{w'})\bigr)\\
&=\Phi_{m_1}\circ\cdots\circ\Phi_{m_{n+1}}(I_H).
\end{aligned}
\]
Applying $\omega$ proves the first equality in
\eqref{eq:observable-reduction-word}.
The second follows from
$\omega(x)=\operatorname{Tr}(\rho_\omega x)$ and
\[
\bigl(\Phi_{m_1}\circ\cdots\circ\Phi_{m_n}\bigr)_*
=\mathcal K_{m_n}\circ\cdots\circ\mathcal K_{m_1}.
\]
Since the cylinder projections span the local commutative
output algebras, these identities determine the restriction
of $\varphi$ to their norm closure.
\end{proof}

\begin{remark}
\label{rem:hybrid}
The Heisenberg-picture maps $\{\Phi_m\}_{m\in O}$ determine precisely
the restriction of $\mathcal G$ to
$\mathcal D_{\mathsf b}\otimes\mathcal A_H$:
\[
\mathcal G\Bigl(\sum_{m\in O}b_m\otimes x_m\Bigr)
=\sum_{m\in O}\Phi_m(x_m),
\qquad x_m\in\mathcal A_H.
\]
In general, however, they do not determine its extension to
$\mathcal A_O\otimes\mathcal A_H$.
The omitted output observables include the off-diagonal blocks
$b_m\mathcal A_O b_{m'}$ for $m\neq m'$ and, when $b_m$ has
rank greater than one, the elements of
$b_m\mathcal A_O b_m$ that are not scalar multiples of $b_m$.
Equation~\eqref{eq:lift-offdiagonal-output} exhibits generators
with the same reduction and different values on the off-diagonal
blocks.

At the level of states, the operational output law is the
restriction of the full output marginal described in
Theorem~\ref{thm:observable-marginal-fcs} to the quasi-local
algebra generated by the copies of $\mathcal D_{\mathsf b}$.
This restriction of states is distinct from $R_{\mathsf b}$,
which maps transition expectations to instruments.
\end{remark}

The reduced instrument determines the hidden channel:
\begin{equation}
\label{eq:reduction-gives-channel}
\mathcal E=\sum_{m\in O}\Phi_m,
\qquad
\mathcal E_*=\sum_{m\in O}\mathcal K_m.
\end{equation}

\subsection{Every operational HQMM admits a Mealy lift}
\label{subsec:lifting}

Throughout this subsection, fix an operational HQMM
$(\mathcal H,\rho_0,\{\mathcal K_m\}_{m\in O})$
over a finite alphabet $O$, with $\mathcal H$ finite-dimensional.
Set $\mathcal H_H=\mathcal H$ and
$\omega_0(a)=\operatorname{Tr}(\rho_0a)$; as usual,
$\mathcal A_H=\mathcal B(\mathcal H_H)$ and
$d_H=\dim\mathcal H_H$.

To construct its lifts, choose $\mathcal H_O$ with an
orthonormal basis $\{|m\rangle\}_{m\in O}$, so that
$\mathcal A_O=\mathcal B(\mathcal H_O)$ and $d_O=|O|$, and
specialize the readout \eqref{eq:readout} to
$b_m=|m\rangle\langle m|$.
Fix a unit vector $|\chi\rangle\in\mathcal H_H$.

\begin{theorem}
\label{thm:mealy-lift}
Choose Kraus decompositions as in \eqref{eq:hqmm-kraus},
\[
\mathcal K_m(\rho)
=\sum_\alpha
L_{m,\alpha}\rho L_{m,\alpha}^*,
\qquad m\in O,
\]
and define
\begin{equation}
\label{eq:general-lift}
V_{m,\alpha}:\mathcal H_H
\longrightarrow
\mathcal H_H\otimes\mathcal H_O\otimes\mathcal H_H,
\qquad
V_{m,\alpha}\psi
=|\chi\rangle\otimes|m\rangle\otimes L_{m,\alpha}\psi.
\end{equation}
Then
\begin{equation}
\label{eq:general-lift-T}
\mathcal F_{\mathrm E}^{\mathrm{inc}}(Z)
=\sum_{m\in O}\sum_\alpha
V_{m,\alpha}^*ZV_{m,\alpha},
\qquad
Z\in\mathcal A_H\otimes\mathcal A_O\otimes\mathcal A_H,
\end{equation}
is a Mealy transition expectation satisfying
\[
R_{\mathsf b}(\mathcal F_{\mathrm E}^{\mathrm{inc}})
=\{\mathcal K_m\}_{m\in O}.
\]
Together with the boundary state $\omega_0$, it generates a Mealy
process whose word probabilities for $\mathsf b$ coincide with
those of the given operational HQMM.

Consequently, on the Mealy transition expectations over
$(\mathcal A_H,\mathcal A_O)$ and for the rank-one readout
$\mathsf b$, the map $R_{\mathsf b}$ is surjective onto the set of
finite-outcome quantum instruments on $\mathcal A_H$ indexed
by $O$.
\end{theorem}

\begin{proof}
Complete positivity follows from \eqref{eq:choi-kraus}.
By \eqref{eq:hqmm-kraus-completeness},
\[
\sum_{m\in O}\sum_\alpha
V_{m,\alpha}^*V_{m,\alpha}
=
\sum_{m\in O}\sum_\alpha
L_{m,\alpha}^*L_{m,\alpha}
=I_H .
\]
Hence $\mathcal F_{\mathrm E}^{\mathrm{inc}}$ is unital.
For $\ell\in O$ and $x\in\mathcal A_H$,
the rank-one projection $b_\ell=|\ell\rangle\langle\ell|$
gives
\[
\mathcal F_{\mathrm E}^{\mathrm{inc}}
(I_H\otimes b_\ell\otimes x)
=\sum_\alpha
L_{\ell,\alpha}^*xL_{\ell,\alpha}.
\]
The map on the right is the Heisenberg dual of
$\mathcal K_\ell$.
Passing to Schr\"odinger duals proves the asserted reduction.
The equality of word probabilities follows from
Theorem~\ref{thm:observable-reduction}.
\end{proof}
We call \eqref{eq:general-lift-T} the \emph{incoherent erasing
lift}: the adjoint actions of the $V_{m,\alpha}$ are summed
separately, so for every $b\in\mathcal A_O$ only the diagonal
coefficients $\langle m|b|m\rangle$ contribute.

This construction requires no additional restriction
on the Kraus ranks of the operations or on $d_O$
relative to $d_H$.

\begin{remark}
\label{rem:lift-no-obstruction}
For any lift of the prescribed instrument, the reduced
Heisenberg maps $\Phi_m$ are the duals of $\mathcal K_m$.
With the rank-one readout fixed above, an isometric lift
on the same hidden and output spaces exists if and only
if each $\Phi_m$ admits a Kraus representation with
at most $d_H$ operators.

Indeed, for any such lift, \eqref{eq:block-reduction} gives
\[
\Phi_m(x)
=\sum_{i=1}^{d_H}W_{i,m}^*xW_{i,m},
\qquad m\in O.
\]
Conversely, choose such Kraus families and complete them
with zero operators to obtain $d_H$ operators for each
$m\in O$. For an orthonormal basis
$\{|i\rangle\}_{i=1}^{d_H}$ of $\mathcal H_H$, define
\[
W\psi
=\sum_{i=1}^{d_H}\sum_{m\in O}
|i\rangle\otimes|m\rangle\otimes W_{i,m}\psi.
\]
Then
\[
W^*W
=\sum_{m\in O}\sum_{i=1}^{d_H}W_{i,m}^*W_{i,m}
=\sum_{m\in O}\Phi_m(I_H)
=I_H.
\]
Thus $W$ is an isometry, and \eqref{eq:block-reduction}
shows that $\operatorname{Ad}_W$ has the prescribed
observable reduction.

\end{remark}

We next establish nonuniqueness of the observable reduction
already among isometric transition expectations.
For the remainder of this subsection, assume that each
operation admits one Kraus operator, and fix
$A_m\in\mathcal A_H$ such that
\begin{equation}
\label{eq:rank-one-instrument}
\mathcal K_m(\rho)=A_m\rho A_m^*,
\qquad m\in O,
\qquad
\sum_{m\in O}A_m^*A_m=I_H.
\end{equation}
This one-Kraus assumption is used only for the explicit coherent
isometric lifts constructed below; it is not needed for the
surjectivity theorem.
\begin{definition}[The erasing and site-storing lifts]
\label{def:two-lifts}
For the site-storing construction, assume $d_O\leq d_H$
and choose an orthonormal family
$\{|e_m\rangle\}_{m\in O}$ in $\mathcal H_H$.
Define
\begin{equation}
\label{eq:two-lifts}
\begin{aligned}
W_{\mathrm E}\psi
&=\sum_{m\in O}
|\chi\rangle\otimes|m\rangle\otimes A_m\psi,\\
W_{\mathrm S}\psi
&=\sum_{m\in O}
|e_m\rangle\otimes|m\rangle\otimes A_m\psi.
\end{aligned}
\end{equation}
The maps
\[
\mathcal F_{\mathrm E}=\operatorname{Ad}_{W_{\mathrm E}},
\qquad
\mathcal F_{\mathrm S}=\operatorname{Ad}_{W_{\mathrm S}},
\]
that is,
\[
\mathcal F_{\mathrm E}(Z)=W_{\mathrm E}^*ZW_{\mathrm E},
\qquad
\mathcal F_{\mathrm S}(Z)=W_{\mathrm S}^*ZW_{\mathrm S},
\]
are called the \emph{erasing lift} and the
\emph{site-storing lift}, respectively.
They are isometric Mealy transition expectations with
the prescribed reduction, as verified in
Lemma~\ref{lem:two-lifts-reduction}.
\end{definition}

Statements concerning the site-storing lift are understood
under the assumption $d_O\leq d_H$.
Orthogonality of the vectors $|e_m\rangle$ ensures that
distinct output labels have orthogonal hidden-site records;
their unit norms alone suffice for the following lemma.

\begin{lemma}
\label{lem:two-lifts-reduction}
The operators $W_{\mathrm E}$ and $W_{\mathrm S}$ are
isometries, and
\[
R_{\mathsf b}(\mathcal F_{\mathrm E})
=R_{\mathsf b}(\mathcal F_{\mathrm S})
=\{\mathcal K_m\}_{m\in O}.
\]
For any boundary state $\omega$, the Mealy processes
generated by $(\omega,\mathcal F_{\mathrm E})$ and
$(\omega,\mathcal F_{\mathrm S})$ have the same word
probabilities for the readout $\mathsf b$.
\end{lemma}

\begin{proof}
Orthogonality of the output basis gives
\[
W_{\mathrm E}^*W_{\mathrm E}
=\sum_{m\in O}\langle\chi|\chi\rangle A_m^*A_m,
\qquad
W_{\mathrm S}^*W_{\mathrm S}
=\sum_{m\in O}\langle e_m|e_m\rangle A_m^*A_m.
\]
Since $|\chi\rangle$ and the $|e_m\rangle$ are unit vectors,
\eqref{eq:rank-one-instrument} implies
\[
W_{\mathrm E}^*W_{\mathrm E}
=W_{\mathrm S}^*W_{\mathrm S}=I_H.
\]
Their adjoint actions are therefore completely positive
and unital.

For $\ell\in O$ and $x\in\mathcal A_H$, the identity
\[
\langle m|b_\ell|n\rangle
=\delta_{m,\ell}\delta_{n,\ell}
\]
yields
\[
\mathcal F_{\mathrm E}(I_H\otimes b_\ell\otimes x)
=A_\ell^*xA_\ell
=\mathcal F_{\mathrm S}(I_H\otimes b_\ell\otimes x).
\]
The map $x\mapsto A_\ell^*xA_\ell$ is the Heisenberg
dual of $\mathcal K_\ell$.
Passing to Schr\"odinger duals proves the equality of
the observable reductions.
The assertion about word probabilities follows from
Theorem~\ref{thm:observable-reduction}.
\end{proof}

The lifts $\mathcal F_{\mathrm E}$ and
$\mathcal F_{\mathrm S}$ are called \emph{coherent}
because the sums defining their isometries are formed
before taking the adjoint action.
For $a,x\in\mathcal A_H$ and $b\in\mathcal A_O$,
\begin{equation}
\label{eq:coherent-lift-slices}
\begin{aligned}
\mathcal F_{\mathrm E}(a\otimes b\otimes x)
&=\langle\chi|a|\chi\rangle
\sum_{m,n\in O}
\langle m|b|n\rangle A_m^*xA_n,\\
\mathcal F_{\mathrm S}(a\otimes b\otimes x)
&=\sum_{m,n\in O}
\langle e_m|a|e_n\rangle
\langle m|b|n\rangle A_m^*xA_n.
\end{aligned}
\end{equation}

In particular, for $u,v\in O$ with $u\neq v$
and $x\in\mathcal A_H$,
\begin{equation}
\label{eq:lift-offdiagonal-output}
\begin{aligned}
\mathcal F_{\mathrm E}
(I_H\otimes|u\rangle\langle v|\otimes x)
&=A_u^*xA_v,\\
\mathcal F_{\mathrm S}
(I_H\otimes|u\rangle\langle v|\otimes x)
&=0,\\
\mathcal F_{\mathrm E}^{\mathrm{inc}}
(I_H\otimes|u\rangle\langle v|\otimes x)
&=0.
\end{aligned}
\end{equation}
Nevertheless,
\[
\mathcal F_{\mathrm S}
(a\otimes|u\rangle\langle v|\otimes x)
=\langle e_u|a|e_v\rangle A_u^*xA_v,
\]
which need not vanish for a general hidden observable $a$.

\begin{proposition}
\label{prop:two-lifts-hidden}
Let $\varphi^{\mathrm E}$ and $\varphi^{\mathrm S}$ be
the processes generated by the lifts of
Definition~\ref{def:two-lifts} and the boundary state
$\omega_0$, and denote the slot maps \eqref{eq:slot-maps}
and boundary maps \eqref{eq:mealy-T0} of these lifts by
$\mathcal E_a^{\mathrm E}$, $\mathcal E_a^{\mathrm S}$ and
$\mathcal F_\partial^{\mathrm E}$,
$\mathcal F_\partial^{\mathrm S}$, respectively.
For all $a,x\in\mathcal A_H$,
\begin{align}
\label{eq:lift-E-slot}
\mathcal E_a^{\mathrm E}(x)
&=\langle\chi|a|\chi\rangle\,\mathcal E(x),\\
\label{eq:lift-S-slot}
\mathcal E_a^{\mathrm S}(x)
&=\sum_{m\in O}
\langle e_m|a|e_m\rangle A_m^*xA_m,
\end{align}
where
\[
\mathcal E(x)=\sum_{m\in O}A_m^*xA_m
\]
is the hidden channel of both lifts, and
\[
\mathcal F_\partial^{\mathrm E}(a)
=\langle\chi|a|\chi\rangle I_H,
\qquad
\mathcal F_\partial^{\mathrm S}(a)
=\sum_{m\in O}\langle e_m|a|e_m\rangle A_m^*A_m.
\]
Moreover:
\begin{enumerate}
\item[(i)]
The hidden marginal of $\varphi^{\mathrm E}$ is the
product state with one-site density operator
$|\chi\rangle\langle\chi|$. Explicitly, for $r\geq1$,
$0\leq k_1<\cdots<k_r$, and
$a_1,\ldots,a_r\in\mathcal A_H$,
\[
\varphi^{\mathrm E}
\bigl(a_1^{(k_1)}\cdots a_r^{(k_r)}\bigr)
=\prod_{j=1}^r\langle\chi|a_j|\chi\rangle.
\]
Every one-site hidden marginal therefore has
von Neumann entropy $0$.

\item[(ii)]
The hidden marginal of $\varphi^{\mathrm S}$ at site $0$
has density operator
\[
\sigma_0^{\mathrm S}
=\sum_{m\in O}p_m|e_m\rangle\langle e_m|,
\qquad
p_m=\operatorname{Tr}\bigl(\mathcal K_m(\rho_0)\bigr).
\]
Consequently,
\[
S(\sigma_0^{\mathrm S})
=H(p)
=-\sum_{m\in O}p_m\log_2p_m,
\]
where $0\log_2 0=0$ and $p$ is the distribution
of the first emitted symbol.
\end{enumerate}
\end{proposition}

\begin{proof}
Setting $b=I_O$ in \eqref{eq:coherent-lift-slices}
gives \eqref{eq:lift-E-slot} and \eqref{eq:lift-S-slot}.
Taking $a=I_H$ in these identities yields
\[
\mathcal E_{I_H}^{\mathrm E}
=\mathcal E_{I_H}^{\mathrm S}
=\mathcal E,
\]
since $|\chi\rangle$ and the $|e_m\rangle$ are unit
vectors; by \eqref{eq:hidden-channel}, $\mathcal E$ is
therefore the hidden channel of both lifts.
Moreover, \eqref{eq:rank-one-instrument} gives
$\mathcal E(I_H)=I_H$.
Evaluating \eqref{eq:lift-E-slot} and
\eqref{eq:lift-S-slot} at $x=I_H$ therefore gives
the boundary maps.

For (i), substitute \eqref{eq:lift-E-slot} into
\eqref{eq:hidden-correlation}.
Each inserted observable $a_j$ contributes the scalar
$\langle\chi|a_j|\chi\rangle$.
The remaining powers of $\mathcal E$ send $I_H$ to
$I_H$, and $\omega_0(I_H)=1$.
This proves the product formula.
Each one-site density operator is a rank-one projection,
so its entropy is zero.

For (ii), \eqref{eq:hidden-one-point} at $k=0$ gives,
for every $a\in\mathcal A_H$,
\[
\begin{aligned}
\varphi^{\mathrm S}(a^{(0)})
&=\omega_0\bigl(\mathcal F_\partial^{\mathrm S}(a)\bigr)\\
&=\sum_{m\in O}
\langle e_m|a|e_m\rangle\,\omega_0(A_m^*A_m)\\
&=\operatorname{Tr}(\sigma_0^{\mathrm S}a),
\end{aligned}
\]
because
\[
\omega_0(A_m^*A_m)
=\operatorname{Tr}(A_m\rho_0A_m^*)
=p_m.
\]
Since the vectors $|e_m\rangle$ are orthonormal,
the eigenvalues of $\sigma_0^{\mathrm S}$, counted
with multiplicity, are the $p_m$ together with
$d_H-d_O$ additional zeros.
The entropy formula follows.
\end{proof}

\begin{remark}
\label{rem:lift-interpretation}
In the erasing lift, each emitted hidden site is prepared in the
fixed state $|\chi\rangle\langle\chi|$; in the site-storing lift, the
orthonormal vectors $|e_m\rangle$ furnish a perfectly distinguishable
record of the corresponding output symbol. By
Corollary~\ref{cor:factorization-channel}, the block span of every
isometric lift of the instrument \eqref{eq:rank-one-instrument} is the
span of the Kraus operators of its hidden channel
$\mathcal E(x)=\sum_mA_m^*xA_m$, namely
$\operatorname{span}\{A_m:m\in O\}$; for the two lifts above this is
also visible from their blocks
$(W_{\mathrm E})_{i,m}=\langle i|\chi\rangle A_m$ and
$(W_{\mathrm S})_{i,m}=\langle i|e_m\rangle A_m$. Hence, by
Theorem~\ref{thm:isometric-moore-characterization}, every isometric
lift admits a Moore factorization precisely when
$\dim\operatorname{span}\{A_m:m\in O\}\leq d_H$, whatever the number of
outcomes. When the site-storing lift exists,
$\dim\operatorname{span}\{A_m:m\in O\}\leq|O|=d_O\leq d_H$, so both
lifts admit Moore factorizations; this holds for the two cluster lifts
of Section~\ref{sec:cluster}, where $d_H=2$. The nonuniqueness
established below therefore arises already within the class of
Moore-factorizable generators.
\end{remark}

\begin{corollary}
\label{cor:channel-not-chain}
The observable reduction determines the hidden channel, but it need
not determine the hidden transition expectation $E_H^{\mathcal F}$,
nor, even with the boundary state fixed, the associated hidden
marginal state.

More precisely, assume \eqref{eq:rank-one-instrument}, let
$d_O\leq d_H$, and suppose that
\[
p_m=\operatorname{Tr}\bigl(\mathcal K_m(\rho_0)\bigr)
\]
satisfies $p_m>0$ for at least two $m\in O$.
Then the two lifts of Definition~\ref{def:two-lifts} have the same
observable reduction and, with boundary state $\omega_0$, the same
word probabilities, while
\[
S\bigl(\varphi^{\mathrm E}|_{\mathcal A_0^H}\bigr)=0,
\qquad
S\bigl(\varphi^{\mathrm S}|_{\mathcal A_0^H}\bigr)
=H(p)>0.
\]
Their hidden marginal states are therefore distinct; since both lifts
carry the same boundary state, their hidden transition expectations
are distinct as well.

Furthermore, the two processes are not related by a factorwise inner
automorphism: there is no automorphism $\alpha$ of
$\mathcal A_\infty$ with
\[
\alpha|_{\mathcal A_k^H}=\operatorname{Ad}_{u_k^H}
\quad(k\geq0),
\qquad
\alpha|_{\mathcal A_\ell^O}=\operatorname{Ad}_{u_\ell^O}
\quad(\ell\geq1),
\]
for unitaries $u_k^H\in\mathcal A_k^H$ and
$u_\ell^O\in\mathcal A_\ell^O$, such that
$\varphi^{\mathrm S}=\varphi^{\mathrm E}\circ\alpha$.
\end{corollary}

\begin{proof}
By \eqref{eq:reduction-gives-channel},
$\mathcal E=\sum_{m\in O}\Phi_m$ and
$\mathcal E_*=\sum_{m\in O}\mathcal K_m$, so the observable reduction
determines the hidden channel.

Lemma~\ref{lem:two-lifts-reduction} gives both
$R_{\mathsf b}(\mathcal F_{\mathrm E})
=R_{\mathsf b}(\mathcal F_{\mathrm S})$
and, for the common boundary state $\omega_0$, the equality of the
word probabilities for $\mathsf b$.

By Proposition~\ref{prop:two-lifts-hidden}(i) the one-site hidden
marginal of $\varphi^{\mathrm E}$ is the pure state
$|\chi\rangle\langle\chi|$, whence
$S(\varphi^{\mathrm E}|_{\mathcal A_0^H})=0$, and by
Proposition~\ref{prop:two-lifts-hidden}(ii)
\[
S\bigl(\varphi^{\mathrm S}|_{\mathcal A_0^H}\bigr)
=S\bigl(\sigma_0^{\mathrm S}\bigr)
=H(p).
\]
By assumption $p_m>0$ for at least two $m\in O$, so $p$ is not
concentrated at a single outcome and $H(p)>0$.
The two one-site hidden marginal states are therefore distinct.

Suppose the two hidden transition expectations were equal. By
Theorem~\ref{thm:hidden-marginal} the hidden marginal of a Mealy
process is the quantum Markov chain generated by the boundary state
and that transition expectation; with the common boundary state
$\omega_0$ the two lifts would then have the same hidden marginal,
contradicting the preceding display. Hence their hidden transition
expectations differ.

Finally, let $\alpha$ be as in the statement and suppose
$\varphi^{\mathrm S}=\varphi^{\mathrm E}\circ\alpha$. Restriction to
$\mathcal A_0^H$ gives
$\varphi^{\mathrm S}|_{\mathcal A_0^H}
=\varphi^{\mathrm E}|_{\mathcal A_0^H}\circ\operatorname{Ad}_{u_0^H}$,
so the two one-site density operators are unitarily equivalent and
have equal entropies, contradicting the values $0$ and $H(p)>0$
obtained above.
\end{proof}

\section{Faithfulness to the Classical Mealy Model}
\label{sec:faithfulness}

This section establishes that the Mealy construction extends the full classical model, including its joint hidden--output process, and not merely the induced output process.

\subsection{The two roles of the hidden variable}
\label{subsec:two-roles}

Let \((\pi,p(\cdot,\cdot\mid\cdot))\) be a classical Mealy hidden Markov model in the sense of Definition~\ref{def:classical-mealy}. Denote its hidden process by \((X_n)_{n\geq0}\), its output process by \((Y_n)_{n\geq1}\), and its associated commutative transition expectation by \(\mathcal F_{\mathrm{cl}}\), as in \eqref{eq:classical-mealy-te}.

The constructions of Theorems~\ref{thm:mealy-existence} and~\ref{thm:hidden-marginal} use only that the algebras involved are finite-dimensional unital \(C^*\)-algebras. Since \(\mathcal F_{\mathrm{cl}}\) is completely positive and unital (Proposition~\ref{prop:classical-mealy-te}), it is a Mealy transition expectation on \((C(S),C(O))\), and these theorems give its quasi-local state and its hidden marginal.

For \(f,h\in C(S)\), define the classical slot maps by
\[
\mathcal E_f^{\mathrm{cl}}(h)
:=\mathcal F_{\mathrm{cl}}
   (f\otimes\mathbf 1_O\otimes h),
\qquad
\mathcal E^{\mathrm{cl}}
:=\mathcal E_{\mathbf 1_S}^{\mathrm{cl}}.
\]

\begin{lemma}
\label{lem:copying}
For every \(f,h\in C(S)\),
\begin{equation}
\label{eq:copying}
\mathcal E_f^{\mathrm{cl}}(h)
=
M_f\bigl(\mathcal E^{\mathrm{cl}}(h)\bigr),
\end{equation}
where \(M_f:C(S)\to C(S)\) denotes multiplication by \(f\),
\[
(M_f g)(i)=f(i)g(i),
\]
and
\[
\bigl(\mathcal E^{\mathrm{cl}}h\bigr)(i)
=
\sum_{m\in O}\sum_{j\in S}
p(j,m\mid i)\,h(j).
\]
\end{lemma}

\begin{proof}
By \eqref{eq:classical-mealy-te},
\(\bigl(\mathcal E_f^{\mathrm{cl}}h\bigr)(i)
=f(i)\sum_{m\in O}\sum_{j\in S}p(j,m\mid i)\,h(j)
=\bigl(M_f\mathcal E^{\mathrm{cl}}h\bigr)(i)\) for every \(i\in S\).
\end{proof}

The identity \eqref{eq:copying} expresses the two roles of the classical hidden variable. For a current state \(i\), the factor \(f(i)\) records the observable at the present hidden site, whereas \(\bigl(\mathcal E^{\mathrm{cl}}h\bigr)(i)\) is the conditional expectation of the next-site observable \(h\) given \(X_n=i\). Thus the current hidden state serves simultaneously as the site label and as the conditioning variable of the transition.

The following proposition shows that, on a full matrix algebra, an
identity boundary map forces the hidden channel to have scalar range.

\begin{proposition}
\label{prop:copying-scalar}
Let \(\mathcal A_H=\mathcal B(\mathcal H_H)\), and let
\[
E:\mathcal A_H\otimes\mathcal A_H\longrightarrow\mathcal A_H
\]
be completely positive and unital, with \(E(a\otimes I_H)=a\) for every
\(a\in\mathcal A_H\). Then there is a state \(\tau\) on \(\mathcal A_H\)
such that
\[
E(a\otimes x)=\tau(x)a,\qquad a,x\in\mathcal A_H,
\]
that is, \(E=\operatorname{id}_{\mathcal A_H}\otimes\tau\); in
particular \(\mathcal E(x)=E(I_H\otimes x)=\tau(x)I_H\). Conversely,
every such map satisfies \(E(a\otimes I_H)=a\).
\end{proposition}

\begin{proof}
On \(\mathcal A_H\otimes I_H\) the map \(E\) is the identity, hence a
\(*\)-homomorphism, so \(\mathcal A_H\otimes I_H\) lies in the
multiplicative domain of \(E\) \cite{Choi1974}:
\(E((a\otimes I_H)Z)=aE(Z)\) and \(E(Z(a\otimes I_H))=E(Z)a\) for all
\(Z\). With \(Z=I_H\otimes x\) this gives
\(a\,\mathcal E(x)=E(a\otimes x)=\mathcal E(x)\,a\). Hence
\(\mathcal E(x)\) lies in the centre \(\mathbb C I_H\), say
\(\mathcal E(x)=\tau(x)I_H\), and \(\tau\) is positive and unital,
hence a state. Thus \(E(a\otimes x)=a\,\mathcal E(x)=\tau(x)a\). The
converse is immediate.
\end{proof}

Each copying identity, \(E(a\otimes x)=a\,\mathcal E(x)\) or its
symmetrized form
\(E(a\otimes x)=\tfrac12\bigl(a\mathcal E(x)+\mathcal E(x)a\bigr)\), gives
\(E(a\otimes I_H)=a\) at \(x=I_H\), since \(\mathcal E(I_H)=I_H\).
Proposition~\ref{prop:copying-scalar} is related to the no-cloning and
no-broadcasting theorems \cite{BarnumEtAl1996,Lindblad1999}.

\begin{remark}
\label{rem:copying-vs-erasing}
The scalar conclusion uses
\(Z(\mathcal B(\mathcal H_H))=\mathbb C I_H\).
For \(C(S)\), centrality imposes no restriction, since the algebra
is commutative.

In the Mealy setting, \(E_H^{\mathcal F}(a\otimes I_H)=\mathcal F_\partial(a)\);
if \(\mathcal F_\partial=\operatorname{id}\), Proposition~\ref{prop:copying-scalar}
applies to \(E=E_H^{\mathcal F}\).
The hidden channel has predual
\(\mathcal E_*(\rho)=\operatorname{Tr}(\rho)\rho_\tau\),
where \(\rho_\tau\) is the density operator of \(\tau\).
Moreover, \eqref{eq:hidden-te} and \eqref{eq:mealy-T0} give
\[
\mathcal E_a(x)=\tau(x)a,
\qquad
\mathcal F_\partial(a)=a.
\]
Substitution into \eqref{eq:hidden-correlation} yields
\[
\varphi(a_0^{(0)}\cdots a_n^{(n)})
=\omega(a_0)\prod_{k=1}^{n}\tau(a_k).
\]
Thus the hidden marginal is
\(\omega\otimes\tau\otimes\tau\otimes\cdots\),
and it is stationary precisely when \(\omega=\tau\).

\end{remark}

\begin{theorem}[Faithfulness of the Mealy extension]
\label{thm:faithfulness}
Let \(\varphi_{\mathrm{cl}}\) be the quasi-local state generated by
\((\pi,\mathcal F_{\mathrm{cl}})\). For every \(n\geq1\),
\(f_0,\ldots,f_n\in C(S)\), and \(g_1,\ldots,g_n\in C(O)\),
\begin{equation}
\label{eq:faithfulness}
\begin{aligned}
&\varphi_{\mathrm{cl}}
\bigl(f_0\otimes g_1\otimes f_1\otimes\cdots
      \otimes g_n\otimes f_n\bigr)\\
&\quad=
\sum_{\substack{i_0,\ldots,i_n\in S\\
                m_1,\ldots,m_n\in O}}
\pi(i_0)
\prod_{k=1}^{n}p(i_k,m_k\mid i_{k-1})
\prod_{k=0}^{n}f_k(i_k)
\prod_{k=1}^{n}g_k(m_k).
\end{aligned}
\end{equation}
Hence \(\varphi_{\mathrm{cl}}\) is the joint law of
\((X_0,Y_1,X_1,\ldots)\), with hidden and observable marginals
equal to the laws of \((X_n)_{n\geq0}\) and
\((Y_n)_{n\geq1}\), respectively.
\end{theorem}

\begin{proof}
By \eqref{eq:mealy-T0}, \eqref{eq:classical-mealy-te}, and the
normalization of \(p\),
\[
\mathcal F_{\mathrm{cl},\partial}(f)(i)
=f(i)\sum_{m\in O}\sum_{j\in S}p(j,m\mid i)
=f(i).
\]
Thus the boundary map is the identity on \(C(S)\).

For the elementary tensor in \eqref{eq:faithfulness},
the iteration \eqref{eq:mealy-iteration} takes the form
\[
\begin{aligned}
h_n&=f_n,\\
h_{k-1}(i)
&=f_{k-1}(i)
\sum_{m\in O}\sum_{j\in S}
p(j,m\mid i)\,g_k(m)\,h_k(j),
\qquad k=n,\ldots,1.
\end{aligned}
\]
Its value under \(\varphi_{\mathrm{cl}}\) is \(\pi(h_0)\) by
Theorem~\ref{thm:mealy-existence}. Expanding this recursion
gives \eqref{eq:faithfulness}.

The coefficient
\(\pi(i_0)\prod_{k=1}^{n}p(i_k,m_k\mid i_{k-1})\)
is the probability of the classical trajectory
\[
(i_0,m_1,i_1,\ldots,m_n,i_n).
\]
Since elementary tensors span each finite local algebra,
\eqref{eq:faithfulness} identifies the quasi-local state with
the joint classical law. Setting all \(g_k=\mathbf 1_O\),
respectively all \(f_k=\mathbf 1_S\), gives the hidden and
observable marginals.
\end{proof}

\begin{remark}
\label{rem:classical-vs-quantum-records}
For a commutative memory the observable reduction determines the
classical kernel: by Proposition~\ref{prop:commutative-bijection}, if
$e_j$ is the indicator of $j\in S$, then
\[
p(j,m\mid i)=(\mathcal P_m e_j)(i),
\]
and together with $\pi$ this kernel determines the whole trajectory law
\eqref{eq:classical-trajectory-law}. For instance, with
$\mathcal E^{\mathrm{cl}}=\sum_{m\in O}\mathcal P_m$, the copying
identity \eqref{eq:copying} gives
\[
\mathbb E\bigl[f(X_0)g(X_1)\bigr]
=\pi\bigl(\mathcal E^{\mathrm{cl}}_f(g)\bigr)
=\pi\bigl(f\,\mathcal E^{\mathrm{cl}}(g)\bigr).
\]
Classical hidden correlations are therefore recoverable even when only
the instrument is given, with no tripartite generator.

For a noncommutative memory, a quantum instrument specifies output
probabilities and conditional memory updates, but does not select a
joint state of the unmeasured physical hidden sites at different times;
the Mealy lift supplies that additional state. Specified interventions
on the memory define their own multi-time statistics, which are to be
distinguished from the passive site marginals studied here.
\end{remark}

\section{Different Output Entanglement for the Same Instrument}
\label{sec:cluster}

An operational instrument and an initial memory state determine
the word probabilities and conditional memory states, but need
not determine the quantum state of the outputs before readout.
We illustrate this distinction using the cluster instrument of
\cite{MonrasBeigeWiesner2011}: two Mealy lifts have the same
operational reduction, while one has entangled output marginals
and the other has fully separable output marginals. Entanglement in
hidden Markov models has also been studied in \cite{SS23,AGLS24Q3,SS25}.

Let $\mathcal H_H=\mathcal H_O=\mathbb C^2$, let $X,Y,Z$ denote
the Pauli matrices, with $Y=iXZ$, and put
$|\pm\rangle=(|0\rangle\pm|1\rangle)/\sqrt2$.
Fix an orthonormal basis $\{|\epsilon_0\rangle,|\epsilon_1\rangle\}$
of $\mathcal H_O$ and the readout $\mathsf b=\{b_0,b_1\}$,
$b_s=|\epsilon_s\rangle\langle\epsilon_s|$.
The cluster instrument is $\mathcal K_s(\rho)=K_s\rho K_s^*$, with
Kraus operators \cite[Eq.~(37)]{MonrasBeigeWiesner2011}
\begin{equation}
\label{eq:monras-kraus}
K_s=\frac1{\sqrt2}
\bigl(|0\rangle\langle\epsilon_s|
     +|1\rangle\langle\epsilon_s|Z\bigr),
\qquad s=0,1,
\end{equation}
which satisfy $\sum_sK_s^*K_s=I_H$.
We use the initial memory state $\rho_0=|+\rangle\langle+|$, that is,
the boundary state $\omega_0(x)=\langle+|x|+\rangle$ for the lifts
below, so that
\[
\mathbb P(s_1\cdots s_n)
=\|K_{s_n}\cdots K_{s_1}|+\rangle\|^2.
\]

Choose a unit vector $|\chi\rangle\in\mathcal H_H$ and an orthonormal
basis $\{|f_0\rangle,|f_1\rangle\}$ of the hidden-site space.
Applying Definition~\ref{def:two-lifts} with output basis
$\{|\epsilon_s\rangle\}$, Kraus operators $K_s$ and
$|e_s\rangle=|f_s\rangle$, and writing $W_{\mathrm{cs}}$ for the
erasing lift $W_{\mathrm E}$, we obtain
\begin{equation}
\label{eq:cluster-frame-form}
\begin{aligned}
W_{\mathrm{cs}}\psi
&=\sum_{s=0}^1
|\chi\rangle\otimes|\epsilon_s\rangle\otimes K_s\psi,\\
W_{\mathrm S}\psi
&=\sum_{s=0}^1
|f_s\rangle\otimes|\epsilon_s\rangle\otimes K_s\psi.
\end{aligned}
\end{equation}
By Lemma~\ref{lem:two-lifts-reduction}, both maps are isometries,
and $\mathcal F_{\mathrm{cs}}=\operatorname{Ad}_{W_{\mathrm{cs}}}$ and
$\mathcal F_{\mathrm S}=\operatorname{Ad}_{W_{\mathrm S}}$ satisfy
\[
\mathcal F_{\mathrm{cs}}(I_H\otimes b_s\otimes x)
=\mathcal F_{\mathrm S}(I_H\otimes b_s\otimes x)
=K_s^*xK_s .
\]
Thus they have the same instrument and, for the boundary state
$\omega_0$, the same word probabilities and conditional memory
states. By \eqref{eq:lift-offdiagonal-output}, applied with the output
basis $\{|\epsilon_s\rangle\}$, they differ on the off-diagonal output
blocks, which the instrument does not retain.
Let
\[
|C_N\rangle
=U_{N-1,N}\cdots U_{1,2}|+\rangle^{\otimes N}
\]
be the one-dimensional cluster state \cite{BriegelRaussendorf2001},
where $U_{j,j+1}$ is the controlled-$Z$ gate on neighbouring qubits.
For $\bullet\in\{\mathrm{cs},\mathrm S\}$, write $\rho^\bullet_{[1,n]}$
for the density operator of the first $n$ outputs of the Mealy process
generated by $(\omega_0,\mathcal F_\bullet)$, and
$\Delta_{\mathsf b}(z)=\sum_sb_szb_s$ for the readout dephasing channel.

These marginals are computed by iterating the isometries.
For an isometric generator $\mathcal F=\operatorname{Ad}_W$, put
$\mathcal S=\mathcal H_H\otimes\mathcal H_O$, $V_1=W$ and
$V_{n+1}=(I_{\mathcal S}\otimes V_n)W$. Since
$\mathcal F_\partial(I_H)=I_H$ by \eqref{eq:T0-unital}, induction on
\eqref{eq:mealy-iteration} gives
\begin{equation}
\label{eq:stinespring-finite}
\mathcal F_{[0,n]}(A\otimes I_H)=V_n^*(A\otimes I_H)V_n
\end{equation}
for every $A$ on the first $n$ site--output pairs. Hence, for a
boundary state with density operator $\rho_0$, the marginal on these
pairs is the partial trace of $V_n\rho_0V_n^*$ over the final memory;
$V_n$ applies $W$ successively to the current memory.

\begin{proposition}
\label{prop:storing-separable}
For every $n\geq1$,
\begin{equation}
\label{eq:cluster-identification}
\rho^{\mathrm{cs}}_{[1,n]}
=\operatorname{Tr}_{n+1}|C_{n+1}\rangle\langle C_{n+1}|,
\end{equation}
and
\begin{equation}
\label{eq:storing-diagonal}
\rho^{\mathrm S}_{[1,n]}
=\Delta_{\mathsf b}^{\otimes n}
  \bigl(\rho^{\mathrm{cs}}_{[1,n]}\bigr)
=\sum_{\boldsymbol s\in\{0,1\}^n}\mathbb P(\boldsymbol s)\,
|\epsilon_{\boldsymbol s}\rangle
\langle\epsilon_{\boldsymbol s}|,
\end{equation}
where
$|\epsilon_{\boldsymbol s}\rangle
=|\epsilon_{s_1}\rangle\otimes\cdots\otimes|\epsilon_{s_n}\rangle$.
In particular, every output marginal of the site-storing lift is
fully separable.
\end{proposition}

\begin{proof}
Let $U_{1,2}=I_4-2|11\rangle\langle11|$. Direct contraction gives
\[
(\langle\epsilon_s|\otimes I_H)
U_{1,2}(\psi\otimes|+\rangle)=K_s\psi ,
\]
and summing over $s$,
$W_{\mathrm{cs}}\psi
=|\chi\rangle\otimes U_{1,2}(\psi\otimes|+\rangle)$.
This is the cluster-generation step of
\cite[Fig.~3]{MonrasBeigeWiesner2011} (see also \cite{SchonEtAl2005}):
the current memory is released
as output after coupling to a fresh qubit in $|+\rangle$, which becomes
the next memory. For $W=W_{\mathrm{cs}}$, the vector $V_n|+\rangle$ of
\eqref{eq:stinespring-finite} is therefore, after grouping the hidden
sites first, $|\chi\rangle^{\otimes n}\otimes|C_{n+1}\rangle$, the last
cluster qubit being the final memory. Tracing out the hidden sites and
the final memory proves \eqref{eq:cluster-identification}.

For $W=W_{\mathrm S}$, after the same regrouping,
\[
V_n|+\rangle
=\sum_{\boldsymbol s}
\Bigl(\bigotimes_{j=1}^n|f_{s_j}\rangle\Bigr)
\otimes|\epsilon_{\boldsymbol s}\rangle
\otimes K_{s_n}\cdots K_{s_1}|+\rangle .
\]
Record vectors of distinct words are orthogonal, so tracing out the
hidden sites removes all off-diagonal word terms, and tracing out the
final memory leaves the weights
$\|K_{s_n}\cdots K_{s_1}|+\rangle\|^2=\mathbb P(\boldsymbol s)$.
This proves the second equality in \eqref{eq:storing-diagonal}.
The diagonal entries of $\rho^{\mathrm{cs}}_{[1,n]}$ in the basis
$\{|\epsilon_{\boldsymbol s}\rangle\}$ are the same probabilities,
since the two lifts have the same reduction, and
$\Delta_{\mathsf b}^{\otimes n}$ keeps exactly these entries; this
proves the first equality. A density operator diagonal in a product
basis is a convex combination of product projections, hence fully
separable.
\end{proof}

For a bipartite density operator $\rho$, write
$\mathcal N_{A|B}(\rho)=(\|\rho^{\mathsf T_A}\|_1-1)/2$ for its
negativity, where $\mathsf T_A$ is partial transposition on $A$.

\begin{theorem}
\label{thm:cluster-separation}
For every readout basis, the output marginals of the two lifts
coincide at length one and differ at every length $n\geq2$.
For every $n\geq3$, $\rho^{\mathrm{cs}}_{[1,n]}$ is entangled across
$1\,|\,2\cdots n$, whereas $\rho^{\mathrm S}_{[1,n]}$ is fully
separable. Moreover,
\begin{equation}
\label{eq:cluster-negativity}
\mathcal N_{1|23}\bigl(\rho^{\mathrm{cs}}_{[1,3]}\bigr)=\tfrac12 .
\end{equation}
\end{theorem}

\begin{proof}
Write $g_1=X_1Z_2$, $g_2=Z_1X_2Z_3$, $g_3=Z_2X_3Z_4$ and
$g_4=Z_3X_4$ for the stabilizer generators of $|C_4\rangle$. In
$|C_4\rangle\langle C_4|=2^{-4}\prod_{j=1}^4(I+g_j)$, only
$I,g_1,g_2,g_1g_2$ survive the trace over qubit $4$. With the three-qubit
operators $A=X\otimes Z\otimes I$ and $B=Z\otimes X\otimes Z$ obtained
from $g_1$ and $g_2$, this gives
\begin{equation}
\label{eq:cluster-three-output}
\rho^{\mathrm{cs}}_{[1,3]}
=\tfrac18(I_8+A+B+AB),
\qquad AB=Y_1Y_2Z_3 .
\end{equation}
Partial transposition on qubit $1$ changes only the sign of $AB$.
The commuting self-adjoint unitaries $A,B$ have four joint
eigenspaces, each of dimension two, on which
$(A,B)=(a,b)\in\{-1,1\}^2$. Hence the eigenvalues of
$(\rho^{\mathrm{cs}}_{[1,3]})^{\mathsf T_1}$ are $(1+a+b-ab)/8$,
that is, $\tfrac14$ six times and $-\tfrac14$ twice. This proves
\eqref{eq:cluster-negativity}. For $n>3$, tracing out outputs
$4,\ldots,n$ returns \eqref{eq:cluster-three-output}; separability
across $1\,|\,2\cdots n$ would imply separability of this three-output
marginal across $1|23$, a contradiction.

Further partial traces give $\rho^{\mathrm{cs}}_{[1,1]}=I_2/2$ and
$\rho^{\mathrm{cs}}_{[1,2]}=\tfrac14(I_4+X\otimes Z)$. The first is
fixed by every readout dephasing, so the marginals coincide at length
one. For the second, write $b_s=(I_2+(-1)^sN)/2$ with
$N=\nu_xX+\nu_yY+\nu_zZ$ and $\nu_x^2+\nu_y^2+\nu_z^2=1$. Since
$\Delta_{\mathsf b}(X)=\nu_xN$ and $\Delta_{\mathsf b}(Z)=\nu_zN$,
Proposition~\ref{prop:storing-separable} gives
\[
\rho^{\mathrm S}_{[1,2]}
=\tfrac14(I_4+\nu_x\nu_zN\otimes N).
\]
Equality with $\rho^{\mathrm{cs}}_{[1,2]}$ would require
$X\otimes Z=\nu_x\nu_zN\otimes N$, which is impossible because the two
operator norms are $1$ and $|\nu_x\nu_z|\leq1/2$. By consistency, the
marginals differ at every length $n\geq2$. Full separability of the
site-storing marginals is Proposition~\ref{prop:storing-separable}.
\end{proof}

The two-output cluster marginal is itself separable:
\[
\rho^{\mathrm{cs}}_{[1,2]}
=\tfrac12\bigl(
|+\rangle\langle+|\otimes|0\rangle\langle0|
+|-\rangle\langle-|\otimes|1\rangle\langle1|
\bigr).
\]
Thus the output states first differ at length two, where the difference
is visible before readout: $\langle X\otimes Z\rangle$ on the first two
outputs equals $1$ for $\mathcal F_{\mathrm{cs}}$ and
$\nu_x^2\nu_z^2\leq\tfrac14$ for $\mathcal F_{\mathrm S}$. Their
entanglement first differs at length three. Both lifts are
Moore-factorizable (Remark~\ref{rem:lift-interpretation}).


\section{Conclusion and Perspectives}
\label{sec:conclusion}

We have formulated hidden quantum Markov processes through a single
completely positive unital transition expectation
$\mathcal F:\mathcal A_H\otimes\mathcal A_O\otimes\mathcal A_H
\to\mathcal A_H$, acting on the present hidden system, the emitted
output and the next hidden system, and we have studied two structural
questions about it. For isometric generators, Moore factorization on
the prescribed hidden algebra is characterized by the dimension of the
block span, which is also the minimal intermediate dimension
(Theorem~\ref{thm:isometric-moore-characterization} and
Corollary~\ref{cor:minimal-intermediate-dimension}). For
$d_H,d_O\geq2$ factorizable isometric generators are non-generic
(Theorem~\ref{thm:moore-nongeneric}), and for isometric generators
factorizability depends only on the Kraus rank of the hidden channel
(Corollary~\ref{cor:factorization-channel}). Relative to a fixed
readout, the observable reduction maps the generators onto the
instruments of operational HQMMs and preserves the word probabilities
(Theorems~\ref{thm:observable-reduction} and~\ref{thm:mealy-lift});
since the instrument determines the hidden channel, Moore
factorizability of isometric generators depends only on the
instrument. On commutative algebras both levels
determine the full joint law of the classical Mealy process, hidden
variables included, once the initial distribution is fixed: the Mealy
construction by Theorem~\ref{thm:faithfulness}, the instrument through
its transition kernel (Remark~\ref{rem:classical-vs-quantum-records}).
They differ for a quantum memory: the cluster instrument of
\cite{MonrasBeigeWiesner2011} is the reduction of a lift whose emitted
systems carry the entangled marginals of the cluster state and of a
lift whose emitted marginals are fully separable
(Theorem~\ref{thm:cluster-separation}).

We are currently investigating a classification of the sets of Mealy transition expectations that admit a prescribed reduction, which remains an open problem in general. We are investigating the fibers of observable reduction and the role of Gram matrices of hidden-site and environmental records in characterizing the output states of Mealy lifts.

The statements of Section~\ref{sec:moore-as-mealy} concern generators on a fixed hidden
algebra, not processes. We have not characterized which Mealy
processes admit no factorizing generator at all, nor the minimal $d_H$
generating a given quantum output process, a question related to the
memory-minimal quantum generation of classical processes
\cite{ZonniosBoydBinder2025} and, when the hidden algebra may be
enlarged, to the classical realization questions of
\cite{VanluytenWillemsDeMoor2008}. For non-isometric generators, the
rank bound of Proposition~\ref{prop:rank-obstruction} is only a
necessary condition for Moore factorization; whether it is sufficient
is open.

The quasi-local construction of
Section~\ref{sec:mealy-hqmp} opens the study of ergodicity, mixing,
mean entropy and correlation decay for Mealy processes, and of
filtering and parameter estimation performed directly at the level of
transition expectations.

\section*{Statements and Declarations}

\noindent\textbf{Funding.} No external funding was received for this research.

\medskip
\noindent\textbf{Competing Interests.} The authors declare that they have no relevant financial or non-financial competing interests.

\medskip
\noindent\textbf{Data Availability.} No datasets were generated or analysed during the current study.

\paragraph{Use of AI tools.}
The authors used the AI assistants Claude (Anthropic) and ChatGPT (OpenAI) as aids for language correction, exploratory analysis, and computation. 


\begin{thebibliography}{99}

\bibitem{CholewaGawronGlombKurzyk2017}
M. Cholewa, P. Gawron, P. G{\l}omb, and D. Kurzyk,
``Quantum hidden Markov models based on transition operation matrices,''
\emph{Quantum Information Processing},
\textbf{16} (2017), Article 101.
DOI: \texttt{10.1007/s11128-017-1544-8}.


\bibitem{accardi1974}
L. Accardi,
``Non-commutative Markov chains,''
in \emph{Proceedings of the International School of Mathematical
Physics},
Universit\`a di Camerino, 1974, pp.~268--295.

\bibitem{AccardiFrigerioLewis1982}
L. Accardi, A. Frigerio, and J.~T. Lewis,
``Quantum stochastic processes,''
\emph{Publications of the Research Institute for Mathematical Sciences},
\textbf{18} (1982), no.~1, pp.~97--133.
DOI: \texttt{10.2977/prims/1195184017}.

\bibitem{AccardiSoueidiLuSouissi2024AHP}
L. Accardi, Soueidi E. G., Y.~G. Lu, and A. Souissi,
``Algebraic hidden processes and hidden Markov processes,''
\emph{Infinite Dimensional Analysis, Quantum Probability and Related
Topics},
\textbf{29} (2026), no.~1, Article~2450009.
DOI: \texttt{10.1142/S0219025724500097}.

\bibitem{AGLS24Q}
L. Accardi, Soueidi E. G., Y.~G. Lu, and A. Souissi,
``Hidden quantum Markov processes,''
\emph{Infinite Dimensional Analysis, Quantum Probability and Related
Topics},
\textbf{28} (2025), no.~4, Article~2450007.
DOI: \texttt{10.1142/S0219025724500073}.

\bibitem{ASS20}
L. Accardi, A. Souissi, and Soueidi E. G.,
``Quantum Markov chains: A unification approach,''
\emph{Infinite Dimensional Analysis, Quantum Probability and Related
Topics},
\textbf{23} (2020), no.~2, Article~2050016.
DOI: \texttt{10.1142/S0219025720500162}.

\bibitem{AGLS24Q3}
L. Accardi, A. Souissi, Soueidi E. G., and M. Rhaima,
``Degree of entanglement in entangled hidden Markov models,''
\emph{Chaos, Solitons \& Fractals}, \textbf{196} (2025),
Article~116389.
DOI: \texttt{10.1016/j.chaos.2025.116389}.

\bibitem{AdhikarySrinivasanGordonBoots2020}
S. Adhikary, S. Srinivasan, G. Gordon, and B. Boots,
``Expressiveness and learning of hidden quantum Markov models,''
in \emph{Proceedings of the Twenty Third International Conference on
Artificial Intelligence and Statistics (AISTATS 2020)},
PMLR \textbf{108}, pp.~4151--4161.

\bibitem{AlRasbiClarkBeige2024}
K. Al Rasbi, L.~A. Clark, and A. Beige,
``Quantum physics cannot be captured by classical linear hidden variable
theories even in the absence of entanglement,''
\emph{Frontiers in Physics}, \textbf{12} (2024), Article~1325239.
DOI: \texttt{10.3389/fphy.2024.1325239}.

\bibitem{Anderson1999}
B.~D.~O. Anderson,
``The realization problem for hidden Markov models,''
\emph{Mathematics of Control, Signals, and Systems},
\textbf{12} (1999), no.~1, pp.~80--120.
DOI: \texttt{10.1007/PL00009846}.

\bibitem{BarnumEtAl1996}
H. Barnum, C.~M. Caves, C.~A. Fuchs, R. Jozsa, and B. Schumacher,
``Noncommuting mixed states cannot be broadcast,''
\emph{Physical Review Letters}, \textbf{76} (1996), pp.~2818--2821.

\bibitem{BaumPetrie1966}
L.~E. Baum and T. Petrie,
``Statistical inference for probabilistic functions of finite-state
Markov chains,''
\emph{The Annals of Mathematical Statistics},
\textbf{37} (1966), no.~6, pp.~1554--1563.
DOI: \texttt{10.1214/aoms/1177699147}.

\bibitem{BinderThompsonGu2018}
F.~C. Binder, J. Thompson, and M. Gu,
``Practical unitary simulator for non-Markovian complex processes,''
\emph{Physical Review Letters}, \textbf{120} (2018), 240502.

\bibitem{BriegelRaussendorf2001}
H.~J. Briegel and R. Raussendorf,
``Persistent entanglement in arrays of interacting particles,''
\emph{Physical Review Letters}, \textbf{86} (2001), no.~5,
pp.~910--913.
DOI: \texttt{10.1103/PhysRevLett.86.910}.

\bibitem{CarlylePaz1971}
J.~W. Carlyle and A. Paz,
``Realizations by stochastic finite automata,''
\emph{Journal of Computer and System Sciences},
\textbf{5} (1971), no.~1, pp.~26--40.
DOI: \texttt{10.1016/S0022-0000(71)80005-3}.

\bibitem{Choi1974}
M.-D. Choi,
``A Schwarz inequality for positive linear maps on $C^*$-algebras,''
\emph{Illinois Journal of Mathematics}, \textbf{18} (1974), pp.~565--574.

\bibitem{Choi1975}
M.-D. Choi,
``Completely positive linear maps on complex matrices,''
\emph{Linear Algebra and its Applications},
\textbf{10} (1975), no.~3, pp.~285--290.
DOI: \texttt{10.1016/0024-3795(75)90075-0}.


\bibitem{CrutchfieldYoung1989}
J.~P. Crutchfield and K. Young,
``Inferring statistical complexity,''
\emph{Physical Review Letters}, \textbf{63} (1989), pp.~105--108.

\bibitem{EphraimMerhav2002}
Y. Ephraim and N. Merhav,
``Hidden Markov processes,''
\emph{IEEE Transactions on Information Theory},
\textbf{48} (2002), no.~6, pp.~1518--1569.
DOI: \texttt{10.1109/TIT.2002.1003838}.

\bibitem{FannesNachtergaeleWerner1992}
M. Fannes, B. Nachtergaele, and R.~F. Werner,
``Finitely correlated states on quantum spin chains,''
\emph{Communications in Mathematical Physics},
\textbf{144} (1992), no.~3, pp.~443--490.
DOI: \texttt{10.1007/BF02099178}.

\bibitem{GuWiesnerRieperVedral2012}
M. Gu, K. Wiesner, E. Rieper, and V. Vedral,
``Quantum mechanics can reduce the complexity of classical models,''
\emph{Nature Communications}, \textbf{3} (2012), Article~762.

\bibitem{Jaeger2000}
H. Jaeger,
``Observable operator models for discrete stochastic time series,''
\emph{Neural Computation},
\textbf{12} (2000), no.~6, pp.~1371--1398.
DOI: \texttt{10.1162/089976600300015411}.

\bibitem{Kraus1983}
K. Kraus,
\emph{States, Effects, and Operations:
Fundamental Notions of Quantum Theory},
Lecture Notes in Physics, vol.~190,
Springer-Verlag, Berlin, 1983.
DOI: \texttt{10.1007/3-540-12732-1}.

\bibitem{Lindblad1999}
G. Lindblad,
``A general no-cloning theorem,''
\emph{Letters in Mathematical Physics}, \textbf{47} (1999), pp.~189--196.

\bibitem{LittmanSuttonSingh2001}
M.~L. Littman, R.~S. Sutton, and S. Singh,
``Predictive representations of state,''
in \emph{Advances in Neural Information Processing Systems 14
(NIPS 2001)},
MIT Press, Cambridge, MA, 2002, pp.~1555--1561.

\bibitem{Mityagin2020}
B.~S. Mityagin,
``The zero set of a real analytic function,''
\emph{Mathematical Notes}, \textbf{107} (2020), no.~3, pp.~529--530.
DOI: \texttt{10.1134/S0001434620030189}.

\bibitem{MonrasBeigeWiesner2011}
A. Monr\`as, A. Beige, and K. Wiesner,
``Hidden quantum Markov models and non-adaptive read-out of many-body
states,''
\emph{Applied Mathematical and Computational Sciences},
\textbf{3} (2011), pp.~93--122.
Also available as arXiv:1002.2337 [quant-ph].

\bibitem{NielsenEtAl2003}
M.~A. Nielsen, C.~M. Dawson, J.~L. Dodd, A. Gilchrist, D. Mortimer,
T.~J. Osborne, M.~J. Bremner, A.~W. Harrow, and A. Hines,
``Quantum dynamics as a physical resource,''
\emph{Physical Review A}, \textbf{67} (2003), Article~052301.

\bibitem{ONeillBarlowSafranekBeige2012}
B. O'Neill, T.~M. Barlow, D. \v{S}afr\'anek, and A. Beige,
``Hidden quantum Markov models with one qubit,''
\emph{AIP Conference Proceedings}, \textbf{1479} (2012), p.~667.
Also available as arXiv:1207.4304 [quant-ph].

\bibitem{PerezGarciaEtAl2007}
D. P\'erez-Garc\'{\i}a, F. Verstraete, M.~M. Wolf, and J.~I. Cirac,
``Matrix product state representations,''
\emph{Quantum Information and Computation}, \textbf{7} (2007),
pp.~401--430.

\bibitem{Rabiner1989}
L.~R. Rabiner,
``A tutorial on hidden Markov models and selected applications in
speech recognition,''
\emph{Proceedings of the IEEE},
\textbf{77} (1989), no.~2, pp.~257--286.
DOI: \texttt{10.1109/5.18626}.

\bibitem{SchonEtAl2005}
C. Sch\"on, E. Solano, F. Verstraete, J.~I. Cirac, and M.~M. Wolf,
``Sequential generation of entangled multiqubit states,''
\emph{Physical Review Letters}, \textbf{95} (2005), Article~110503.



\bibitem{SouissiBarhoumi2026}
A. Souissi and A. Barhoumi,
``Causal architecture in hidden quantum Markov models,''
\emph{arXiv preprint},
arXiv:2602.19120, 2026.
DOI: \texttt{10.48550/arXiv.2602.19120}.

\bibitem{SS25}
A. Souissi, F. Mukhamedov, Soueidi E. G., M. Rhaima, and
F. Mukhamedova,
``Entangled hidden elephant random walk model,''
\emph{Chaos, Solitons \& Fractals}, \textbf{186} (2024),
Article~115252.
DOI: \texttt{10.1016/j.chaos.2024.115252}.

\bibitem{SS23}
A. Souissi and Soueidi E. G.,
``Entangled hidden Markov models,''
\emph{Chaos, Solitons \& Fractals}, \textbf{174} (2023),
Article~113804.
DOI: \texttt{10.1016/j.chaos.2023.113804}.

\bibitem{SrinivasanGordonBoots2018}
S. Srinivasan, G. Gordon, and B. Boots,
``Learning hidden quantum Markov models,''
in \emph{Proceedings of the Twenty-First International Conference on
Artificial Intelligence and Statistics (AISTATS 2018)},
PMLR \textbf{84}, pp.~1979--1987.

\bibitem{TamonXie2019}
C. Tamon and W. Xie,
``A note on quantum Markov models,''
\emph{arXiv preprint},
arXiv:1911.01953 [quant-ph], 2019.
DOI: \texttt{10.48550/arXiv.1911.01953}.

\bibitem{ThonJaeger2015}
M. Thon and H. Jaeger,
``Links between multiplicity automata, observable operator models and
predictive state representations: A unified learning framework,''
\emph{Journal of Machine Learning Research},
\textbf{16} (2015), pp.~103--129.

\bibitem{VanluytenWillemsDeMoor2008}
B. Vanluyten, J.~C. Willems, and B. De Moor,
``Equivalence of state representations for hidden Markov models,''
\emph{Systems \& Control Letters},
\textbf{57} (2008), no.~5, pp.~410--419.
DOI: \texttt{10.1016/j.sysconle.2007.10.004}.

\bibitem{Vidyasagar2011Survey}
M. Vidyasagar,
``The complete realization problem for hidden Markov models:
A survey and some new results,''
\emph{Mathematics of Control, Signals, and Systems},
\textbf{23} (2011), nos.~1--3, pp.~1--65.
DOI: \texttt{10.1007/s00498-011-0066-7}.

\bibitem{Vidyasagar2014Book}
M. Vidyasagar,
\emph{Hidden Markov Processes: Theory and Applications to Biology},
Princeton Series in Applied Mathematics,
Princeton University Press, Princeton, NJ, 2014.
DOI: \texttt{10.23943/princeton/9780691133157.001.0001}.

\bibitem{ZonniosBoydBinder2025}
M. Zonnios, A. Boyd, and F.~C. Binder,
``Quantum generation of stochastic processes: spectral invariants and
memory bounds,''
\emph{New Journal of Physics}, \textbf{27} (2025), 064507.
DOI: \texttt{10.1088/1367-2630/addc10}.
Preprint arXiv:2412.12812.

\end{thebibliography}
\end{document}